\documentclass[lettersize,journal]{IEEEtran}
\usepackage{amsmath,amsfonts}
\usepackage{algorithmic}
\usepackage{algorithm}
\usepackage{array}
\usepackage[caption=false,font=small,labelfont=sf,textfont=sf]{subfig}
\usepackage{textcomp}
\usepackage{stfloats}
\usepackage{url}
\usepackage{verbatim}
\usepackage{graphicx}
\usepackage{overpic}
\usepackage{xcolor}
\usepackage{cite}
\begin{document}

\title{Numerical Model of a Multiple-Input-Multiple-Output Distributed Acoustic Sensor with Joint Phase and Birefringence Estimation}

\author{Diane Prato, Mehran Mokhtari Sheramin, Renaud Gabet, Élie Awwad
\thanks{Manuscript created March, 2026. \\
Diane Prato, Mehran Mokthari Sheramin, Renaud Gabet and Élie Awwad are from LTCI Lab, Télécom Paris, Institut Polytechnique de Paris, 19 place Marguerite Perey, 91120 Palaiseau, France. (email: diane.prato@telecom-paris.fr)}}



\maketitle

\begin{abstract}
In this work, we introduce and experimentally validate a numerical model for a Multiple-Input-Multiple-Output Distributed Acoustic Sensing (MIMO-DAS) system that accounts for dynamic perturbations of fiber birefringence and of the common optical phase of the backscattered signal (or polarization-averaged phase, shared by both polarization tributaries). The MIMO-DAS system probes the fiber using polarization-multiplexed constant-power coded sequences that are suited for coexistence of DAS with WDM data transmission over the same fiber. We study the effect of both axisymmetric and anisotropic events on the two quantities. We demonstrate, through numerical simulations and lab experiments, the estimation of effective birefringence magnitude in static conditions, and the joint estimation of common phase and effective birefringence magnitude in the case of dynamic longitudinal strain and anisotropic transverse strain.  This allows for event discrimination and increased sensitivity to disturbances that act transversely on the fiber, since polarization will be responsive to perturbations that break cylindrical symmetry, while the phase will strongly respond to longitudinal strain. 
\end{abstract}

\begin{IEEEkeywords}
Birefringence monitoring, Distributed Optical Fiber Sensing, MIMO-DAS, Orthogonal Golay sequences, Rayleigh scattering, Waveplate model.
\end{IEEEkeywords}

\section{Introduction}
\label{sec:intro}
\IEEEPARstart{N}{owadays}, optical fiber networks are largely deployed for telecommunications all over the world. Taking advantage of this existing fiber infrastructure for monitoring purposes avoids the deployment of a multitude of discrete, dedicated sensors and hence
greatly reduces the logistical cost (installation, energy supply, maintenance). The goal of distributed fiber sensing is therefore to use fibers as sensors for car traffic and train monitoring, surveillance of protected
sites, earthquake monitoring, or structural health monitoring in optical fiber networks \cite{lindsey_fiberoptic_2017,cedilnik_advances_2018,westbrook_enhanced_2023}. Using this available infrastructure to capture, locate and identify events paves the way for improved real-time network
monitoring and the provision of valuable data for a multitude of applications.
One of the most common approaches to distributed strain sensing over a fiber consists of sending light into the fiber and studying its backreflection caused by Rayleigh scattering \cite{lu_distributed_2019,palmieri_distributed_2013}, a phenomenon inherent to optical fibers.
Most of the current systems focus on estimating the common optical phase (common to both polarization tributaries) of the backscattered light (differential-phase Optical Time Domain Reflectometry, $\Delta\phi$-OTDR)\cite{hartog_introduction_2017}, but often neglecting the polarization aspects. Conversely, Polarization-Optical Time Domain Reflectometry (POTDR)\cite{rogers_distributed_2000} takes advantage of polarization properties of the fiber to detect transverse deformations, including bending, twisting, transverses stresses etc.\cite{palmieri_distributed_2013-polar}, mostly in static conditions. However, its application to dynamic monitoring is partly limited by the interrogation method that requires to send several input States of Polarization (SOPs), resulting in a longer measurement time. Moreover, P-OTDR doesn't exploit the common phase of the backscattered signal and is thus unable to detect purely axisymmetric perturbations. 
Instead of analyzing only the common optical phase with conventional $\Delta\phi$-OTDR or the polarization parameters with a P-OTDR, using a multi-parametric dynamic sensing approach allows higher levels of sensitivity and a wider field of applications than that reported in the state of the art. Modern coded polarization-diversity DAS architectures support simultaneous sensing of phase and polarization characteristics throughout a telecom-grade fiber \cite{guerrier_introducing_2020} through the distributed estimation of round-trip Jones matrices along the fiber. However, this approach has mainly been used to mitigate polarization fading. Exploiting the estimated matrices to extract both phase and polarization-related quantities is therefore of interest to detect both isotropic and anisotropic disturbances (which could be for instance bending, or transverse loading on the fiber), and potentially discriminate between the two. 

In this paper, extending our previous work\cite{SPIEPaper}, we develop a numerical model of a MIMO-DAS system accounting for dynamic birefringence perturbations, propose a joint estimation method for phase and birefringence from distributed Jones matrices, and demonstrate the capability of the approach to discriminate between longitudinal and transverse strain events. In section II, we set out the simulation model to describe propagation and backscattering in the fiber, taking into account common phase and linear birefringence, both in static and dynamic conditions. We then introduce  in section III the MIMO-DAS architecture used in our simulations, and the processing steps for distributed parameters extraction: common phase and effective linear birefringence strength. In section IV, we show the ability of the system to provide a static estimation of the effective birefringence magnitude and to detect, localize and discriminate longitudinal and transverse dynamic strains with a mean spatial resolution of $1.3$m. Section V validates the simulated model through lab experiments, and section VI discusses the limitations of the proposed architecture. 

\section{Fiber model}
In this section, we present the waveplate model used to describe propagation and scattering in the fiber, and set the parameters values in static conditions. We then present how the model parameters are modified in the presence of strain. In this work, we only consider standard single mode fibers. 
\subsection{General waveplate model}
 \begin{figure*}[!t]
    \centering
    \includegraphics[width=0.8\linewidth]{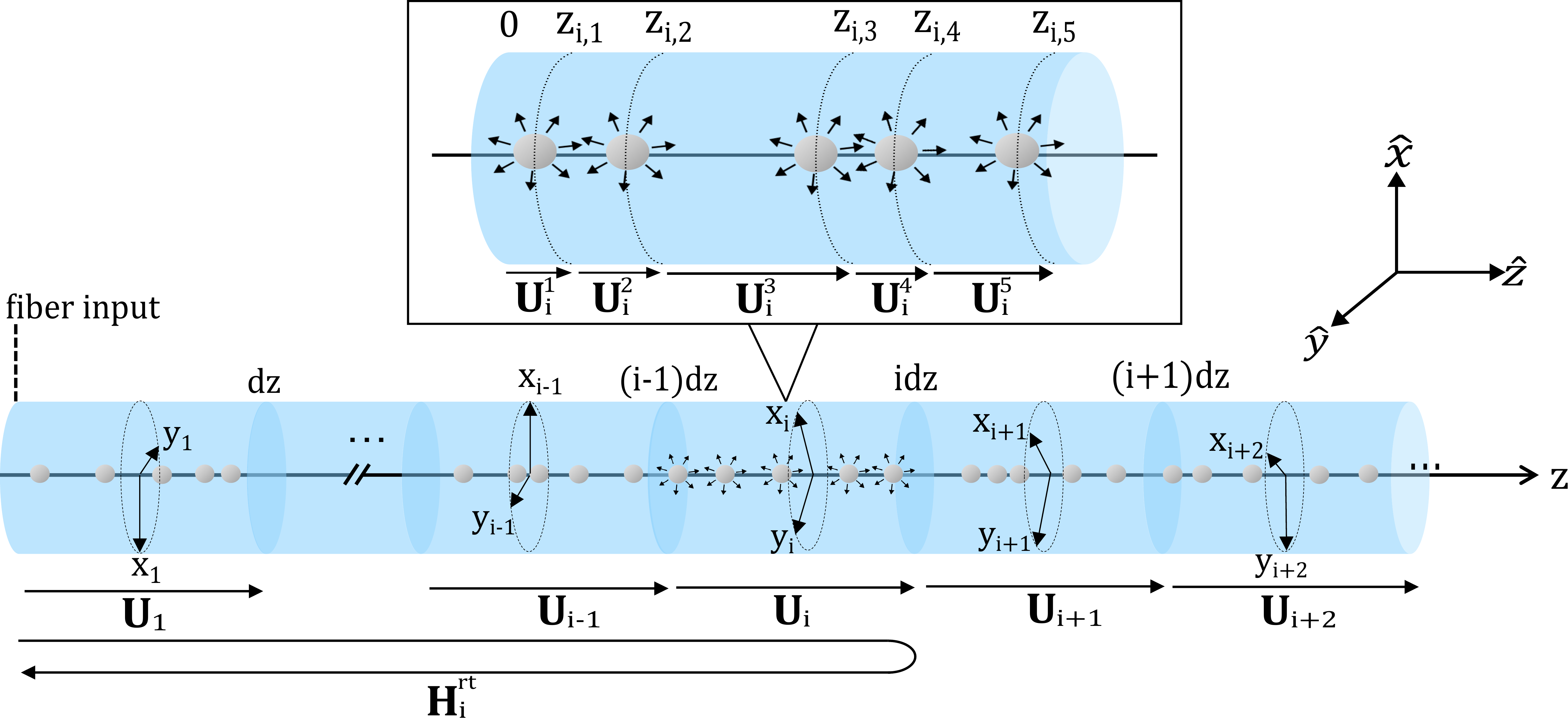}
    \caption{Diagram of the fiber waveplate model (simplified view with 5 scatterers per segment).}
    \label{fig:schema_fibre}
\end{figure*}
\label{sec:fiber_model}
We divide the modeled  sensing fiber into $N_{seg}$ segments of length $dz$ that we assimilate to waveplates according to the waveplate model \cite{pachnicke_fiber-optic_2012, zhao_nonlocal_2025}. The polarization and phase parameters are assumed to be constant accross each segment. Each segment $i$ is described by its forward Jones matrix $\mathbf{U}_i$ representing the propagation: 
\begin{equation}
\begin{split}
    \mathbf{U}_i &= \mathbf{R}^{-1}(\theta_i)\begin{pmatrix} \exp(-j\beta_{x, i}dz) & 0 \\  0 & \exp(-j\beta_{y, i}dz)\end{pmatrix}\mathbf{R}(\theta_i) \\
    &= \exp\left(-j\mathbf{R}^{-1}(\theta_i)\begin{pmatrix} \beta_{x, i} & 0 \\  0 & \beta_{y, i}\end{pmatrix}\mathbf{R}(\theta_i)dz\right)
\end{split}    
\end{equation}
This form accounts for linear birefringence, with the two polarization tributaries having different propagation constants. Defining $\beta_{0,i} = \frac{\beta_{x, i} + \beta_{y, i}}{2}$ and $\Delta\beta_i = \beta_{y, i} - \beta_{x, i}$, we can rewrite $\mathbf{U}_i$ as:
\begin{align}
\begin{split}
    \mathbf{U}_i &=\exp(-j\beta_{0,i}dz)  \\ & \quad \times \mathbf{R}^{-1}(\theta_i)\begin{pmatrix} \exp(j\frac{\Delta\beta_i}{2} dz) & 0 \\  0 & \exp(-j\frac{\Delta\beta_i}{2} dz) \end{pmatrix}\mathbf{R}(\theta_i) \\
    &= \exp\left(-j\left(\beta_{0,i}\mathbf{I}+ \mathbf{R}^{-1}(\theta_i)\begin{pmatrix} \frac{\Delta\beta_i}{2} & 0 \\  0 & \frac{\Delta\beta_i}{2}\end{pmatrix}\mathbf{R}(\theta_i)\right)dz\right)
\end{split}
\end{align} where $\times$ denotes the usual product.
The rotation matrix $\mathbf{R}(\theta_i) = \begin{pmatrix}
    \cos\theta_i
 & -\sin\theta_i \\ \sin\theta_i & \cos\theta_i \end{pmatrix}$ rotates the polarization state into the segment's eigenaxes: $\theta_i$ represents the angle between the segment's fast axis and the $\hat{x}$ axis of the reference frame: it will be referred to as the birefringence orientation. In the convention $\Delta \beta_i > 0$ and given the definition of $\theta_i$, the fast axis of the segment is x, while the slow axis is y. The matrix $\begin{pmatrix} \exp(j\frac{\Delta\beta_i}{2} dz) & 0 \\  0 & \exp(-j\frac{\Delta\beta_i}{2} dz)\end{pmatrix}$ acts as a linear retarder, introducing a phase shift between the two linear polarization tributaries, and the term $\exp(-j\beta_{0,i}dz)$ is the phase term common to both polarizations (referred to as polarization-averaged phase in \cite{mecozzi_sensing_2024}). This propagation matrix only represents linear birefringence (circular birefringence will not be accounted for in round-trip measurements \cite{zhao_nonlocal_2025,chen_distributed_2021}).
 
 \subsection{Rayleigh scattering}
 \label{sec:scattering}

While the matrices $\mathbf{U}_i$ represent propagation in fiber segments, the description of the scattering region needs a finer approach as light is reflected from all scattering centers in the segment and not simply by a mirror at the segment's end.\\ Each fiber segment $i$ contains $N_{scat}$ scattering centers. We consider $N_{scat}=50$ in our model to maintain reasonable computation times. Each scattering center $k \in [1, N_{scat}]$ in segment $i$ has a position $z_{i,k}$ inside the section (its distance from the segment's start) and amplitude $a_{i,k}$. For a realistic description, we distribute the scatterers positions as such: for each scatterer $k$ in segment $i$ is associated a position $z_{i,k} = (k-1).d + \mathcal{U}(0, d)$\cite{liokumovich_fundamentals_2015,masoudi_analysis_2017}  where $d = \frac{dz}{N_{scat}}$ and \( \mathcal{U}(0,d) \) is a realization of a uniform random variable on \([0,d]\). The scatterers amplitudes $a_{i,k}$, $k = [1,..,N_{scat}]$ are defined according to \cite{liokumovich_fundamentals_2015}. In each fiber segment, the incident light is scattered on each scattering center and part of it is backscattered towards the fiber input. To model the backscattering process, we define for each scatterer $k$ in segment $i$ the propagation matrix 
\begin{align}
    \mathbf{U}_i^k &= \exp(-j\beta_{0,i}z_{i,k}) \notag \\ 
    & \quad\times \mathbf{R}^{-1}(\theta_i)\begin{pmatrix} \exp(j\frac{\Delta\beta_i}{2} z_{i,k}) & 0 \\  0 & \exp(-j\frac{\Delta\beta_i}{2} z_{i,k})\end{pmatrix}\mathbf{R}(\theta_i).
\end{align} The total backscattering in segment $i$ is therefore represented by \begin{equation}
    \mathbf{U}_{i, scat}^{rt} = \sum_{k=1}^{N_{scat}}a_{i,k}(\mathbf{U}_i^k)^T\mathbf{U}_i^k. 
\end{equation}The whole round-trip from the fiber input to segment $i$ and back is defined by: 
\begin{align}
    \label{Hroundtrip}
    \mathbf{H}_i^{rt} &= \exp(-\alpha i dz) \notag \\ & \quad \times  \mathbf{U}_1^T\mathbf{U}_2^T...\mathbf{U}_{i-1}^T\biggl[\sum_{k=1}^{N_{scat}}a_{i,k}(\mathbf{U}_i^k)^T\mathbf{U}_i^k\biggr]\mathbf{U}_{i-1}...\mathbf{U}_2\mathbf{U}_1
\end{align}
  where $\exp{(-\alpha i
 dz)}$ accounts for the round-trip attenuation in the fiber with $\alpha = -0.2$dB/km, and the matrices $\mathbf{U}_r$ for $r\in[1, i-1]$ are defined as in section \ref{sec:fiber_model}. Fig. \ref{fig:schema_fibre} illustrates the described fiber model, with $N_{scat} = 5$ for further clarity. 
 
\subsection{Static conditions}
In the absence of events occuring near the fiber, we assume for simplicity (since the absolute static common phase is unexploited in this work) $\beta_{0,i} = \frac{2\pi}{\lambda}\frac{(n_{x,i}+n_{y,i})}{2} = \frac{2\pi n_0}{\lambda} = \beta_0$ to be constant accross the fiber with $n_0 = 1.456$ the mean effective refractive index of the fiber core, and $\beta_{x, i} = \beta_0 - \Delta \beta_i/2$ and $\beta_{y, i} = \beta_0 + \Delta \beta_i/2$.\\
To generate the initial birefringence magnitudes $\Delta \beta_i$ and orientations $\theta_i$ of each segment $i$, we use the model provided by \cite{wai_polarization_1994,wai_polarization_1996}, where the components $\beta_1$ and $\beta_2$ of the Stokes birefringence vector $\begin{pmatrix}
    \beta_{1,i} \\ \beta_{2,i} \\0
\end{pmatrix}$ are independent Langevin processes along the fiber. The parameters of interest, the fiber autocorrelation length and the beat length, are chosen as $L_C = 100$m \cite{damask_polarization_2005}, $L_b = 29.6$m\cite{wuilpart_measurement_2001}. This allows the generation of the birefringence vector at each position $i$, and the birefringence magnitude and orientation: $\Delta \beta_i = \sqrt{\beta_{1,i}^2 + \beta_{2,i}^2}$ and $\theta_i = 0.5 \arctan{\beta_{2,i}/\beta_{1,i}}$ are inserted into the Jones matrices to stay consistent with a Jones waveplate model. \\
In long fibers, this results in the birefringence strength $\Delta \beta$ being Rayleigh distributed in the fiber, with a mean value of $\frac{2\pi}{L_b}$ rad/m \cite{wuilpart_measurement_2001,corsi_beat_1999}. This static birefringence accounts for internal stresses, bends, asymmetries in the core, in static conditions. 
The round-trip matrix $\mathbf{H}_i^{rt}$ can be written as a sum of phase terms multiplied by unitary matrices: 
\begin{align}
    \mathbf{H}_i^{rt} &= \exp(-\alpha i\, dz)
    \bigg[\sum_{l=1}^{i-1}\exp(-2j\beta_{0,l}dz)\bigg] \notag\\
    &\times \sum_{k=1}^{N_{scat}} a_{i,k}\exp(-2j\beta_0 z_{i,k}) \notag\\
    &\times \widehat{\mathbf{U}}_1^T\widehat{\mathbf{U}}_2^T
    \cdots\widehat{\mathbf{U}}_{i-1}^T
    \biggl[(\widehat{\mathbf{U}}_i^k)^T\widehat{\mathbf{U}}_i^k\biggr]
    \widehat{\mathbf{U}}_{i-1}\cdots\widehat{\mathbf{U}}_2\widehat{\mathbf{U}}_1
\end{align} where $\widehat{.}$ denotes the matrix without its common phase term, which is unitary. 

\subsection{Perturbations}
A wide range of disturbances can happen on an optical fiber, which will impact the fiber parameters differently. Two effects need to be taken into account: the fiber elongation, which will modify the scatterers positions, and the photoelastic effect, which will impact the refractive indices \cite{sharpe_springer_2008}. Given the expressions of $\beta_{0,i}$ and $\Delta \beta_i$, these two parameters will be affected by the change in refractive indices, producing an effect on the common phase and the birefringence of perturbed fiber segments. Depending on the nature of the strain event, these parameters will not see the same changes, making it possible to differentiate events up to a certain extent. 
In the following, we will consider 2 types of events: pure longitudinal strain and anisotropic transverse strain.\\

\subsubsection{The photoelastic effect}
\label{photoelasticity}
The photoelastic effect describes the changes in refractive indices caused by mechanical deformations. Assuming the fiber is initially isotropic (this approximation treats static and perturbation-induced birefringence as independent contributions), the photoelastic effect on the fiber for a strain tensor $\begin{pmatrix}
    \varepsilon_x \\ \varepsilon_y \\ \varepsilon_z
\end{pmatrix}$ (neglecting shear strains) induces~\cite{gafsi_analysis_2000}: 
\begin{equation}
\begin{split}
\label{indices_change}
    \delta n_x = - \frac{n_0^3}{2}(p_{11}\varepsilon_x + p_{12}(\varepsilon_y + \varepsilon_z))  \\
     \delta n_y = - \frac{n_0^3}{2}(p_{11}\varepsilon_y + p_{12}(\varepsilon_x + \varepsilon_z)) \\
    \delta n_z =  - \frac{n_0^3}{2}(p_{11}\varepsilon_z + p_{12}(\varepsilon_x + \varepsilon_y)) 
\end{split}
\end{equation}
where $n_0$ is the unperturbed fiber effective refractive index, $p_{11} = 0.121$ and $p_{12} = 0.270$ are the photoelastic coefficients \cite{Bertholds_86} and $x$, $y$, $z$ the directions of the principal strains. From this stems: 
\begin{equation}
\label{diff_n}
    \delta n_y - \delta n_x = -\frac{n_0^3}{2}(p_{11} - p_{12})(\varepsilon_y - \varepsilon_x)
\end{equation}  
\begin{equation}
\label{sum_n}
    \delta n_x + \delta n_y  = -\frac{n_0^3}{2}((p_{11}+p_{12})(\varepsilon_x + \varepsilon_y)+2p_{12}\varepsilon_z)
\end{equation}

Equation~\eqref{diff_n} indicates that birefringence will arise from non-axisymmetric strains when $\varepsilon_x \neq \varepsilon_y$, while Equation \ref{sum_n} shows that the common phase will be sensitive to both transverse and longitudinal strains. Estimating both parameters allows to gain some knowledge on the nature of the strains in the fiber. 
\subsubsection{Phase and birefringence changes}
\label{perturbations}
To model the effect of a perturbation on the fiber (which has already intrinsic birefringence) on a segment $i$, we define the angle $\gamma$ between the  $\hat{x}$ axis of the reference frame and the fast axis of the birefringence induced by the event (for instance, for a bending event, the induced-birefringence fast axis would be normal to the bending radius, while it would be along the loading direction in the case of a transverse load).

We can update the propagation matrix $\mathbf{U}_i$ and the scattering matrices $\mathbf{U}_i^k$ by adding a birefringence matrix to the previous (intrinsic) birefringence matrix to obtain the new matrix $\mathbf{U'}_i$ (respectively $(\mathbf{U}_i^k)'$). This approach is equivalent to summing the birefringence vectors in Stokes space \cite{palmieri_distributed_2013-polar}. The following derivation holds for both matrices $\mathbf{U}_i$ and $\mathbf{U}_i^k$, with $z'$ denoting the updated length $dz$ in the case of $\mathbf{U}_i$, or the updated $z_{i,k}$ in the case of $\mathbf{U}_i^k$. The exact derivation of the quantity $z'$ is explained later when describing the effect of longitudinal strain. The update is performed as such:
\begin{align}
\mathbf{U}_i' &= \exp\Biggl[-j\Biggl(\mathbf{R}^{-1}(\theta_i)
    \begin{pmatrix} \beta_{x,i} & 0 \\ 0 & \beta_{y,i}\end{pmatrix}
    \mathbf{R}(\theta_i) \notag\\
&\qquad \qquad \quad +\mathbf{R}^{-1}(\gamma)
    \begin{pmatrix} \beta_{x,event} & 0 \\ 0 & \beta_{y,event}\end{pmatrix}
    \mathbf{R}(\gamma)\Biggr)z'\Biggr] \notag\\
&= \exp\Biggl[-j\Biggl(\left(\beta_0+
    \frac{\beta_{x,event}+\beta_{y,event}}{2}\right)\mathbf{I} \notag\\
&\qquad \qquad \quad + \mathbf{R}^{-1}(\eta_i)
    \begin{pmatrix} \frac{\Delta\beta_i'}{2} & 0 \\ 0 & -\frac{\Delta\beta_i'}{2}\end{pmatrix}
    \mathbf{R}(\eta_i)\Biggr)z'\Biggr] \notag\\
&= \exp(-j\beta_{0,i}'z')\, \\
& \quad\times \mathbf{R}^{-1}(\eta_i) \notag
    \begin{pmatrix} 
        \exp\!\left(j\tfrac{\Delta\beta_i'}{2} z'\right) & 0 \\ 
        0 & \exp\!\left(-j\tfrac{\Delta\beta_i'}{2} z'\right) 
    \end{pmatrix}
    \mathbf{R}(\eta_i)
\end{align}
where: \begin{equation}
    \beta_{0,i}' = \beta_{0,i}+\frac{\beta_{x,event}+\beta_{y,event}}{2}
\end{equation}
\begin{equation}
    \tan(2\eta_i) = \frac{\Delta\beta_i \sin(2\theta_i) + \Delta\beta_{event} \sin(2\gamma)}{\Delta\beta_i \cos(2\theta_i) + \Delta\beta_{event} \cos(2\gamma)}
\end{equation}
\begin{equation}
\label{resulting_biref}
    \Delta\beta_i' = \sqrt{\Delta\beta_i^2 + \Delta\beta_{event}^2 + 2\Delta\beta_i\Delta\beta_{event} \cos(2(\theta_i - \gamma))}
\end{equation}
with $\Delta\beta_{event} = \beta_{y,event}-\beta_{x,event}  = \frac{2\pi}{\lambda}\frac{\delta n_{y,event} - \delta n_{x,event}}{2}$.

It is clear from Equation~\eqref{resulting_biref} that the resulting birefringence magnitude $\Delta\beta_i'$ is nonlinear in the perturbation, except for a few particular cases: the fast axes of the induced and intrinsic birefringence are aligned or orthogonal, or the event-induced birefringence magnitude $\Delta \beta_{event}$ is dominant compared to the intrinsic one or the opposite. 
Outside of these cases, the response will be nonlinear in the applied strain.\\

\begin{figure*}[!b]
    \centering
    \includegraphics[width=0.8\linewidth]{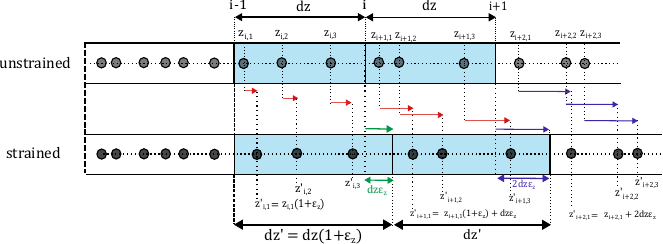}
    \caption{Effect of the application of a pure longitudinal strain $\epsilon_z$: example for a strain affecting 2 fiber segments, $i$ and $i+1$. The beginning of the strained region is fixed, and the fiber is pulled from the right end of the strained region. The applied strain impacts the size of the fiber segments and the scatterers positions: the scatterers inside the strained region furthest from the start will be most affected, and the scatterers further than the end of the strained region are all shifted by the same length.}
    \label{fig:longitudinal_strain}
\end{figure*}
\subsubsection{Longitudinal strain}
\label{longitudinal_strain}
We will assume that a longitudinal strain $\varepsilon_z$ takes place on $p$ segments, between segments $i$ and $i+p$, as if the fiber is being pulled from the end of segment $i+p$, while the starting point of segment $i$ remains fixed. For pure longitudinal strain, $\varepsilon_x = \varepsilon_y = -\nu \varepsilon_z$ according to the Poisson effect, with $\nu = 0.17$ the Poisson's ratio for silica fiber \cite{sharpe_springer_2008}. First, the length $dz$ of the segments under strain is updated to $dz' = dz(1+\varepsilon_z)$. Furthermore, 
the longitudinal strain results in a redistribution of the scatterers positions in the strained region.

 The position of each scatterer $k$ in its respective segment (as described in section \ref{sec:scattering}) is updated as such:
\begin{equation}
    \left\{
    \begin{array}{ll}
        z_{m,k} = z_{m,k} & \mbox{if } m<i \\
        z_{m,k} = z_{m,k} + \varepsilon_z(z_{m,k}+(m-i)dz) & \mbox{if } m \in [i, i+p] \\
        z_{m,k} = z_{m,k} + \varepsilon_zpdz & \mbox{if } m>i+p
    \end{array}
\right.
\end{equation} Fig. \ref{fig:longitudinal_strain} illustrates the scatterers redistribution caused by a longitudinal strain event on two consecutive segments. \\

Another contribution stems from the photoelastic effect: the refractive indices $n_{x,m}$ and $n_{y,m}$ for $m \in [i, i+p]$ are updated according to Equation~\eqref{indices_change} and their sum and difference are computed with Equations~\eqref{sum_n} and \eqref{diff_n}. $(\delta n_{x,m}-\delta n_{y,m})$ being zero for a pure longitudinal strain, no additional birefringence is induced by the event, and the retardance $\Delta\beta_m z$ is scaled by the change in length. Since the preexisting birefringence $\Delta\beta_m$ is generally small, this effect is almost imperceptible in an SSMF (standard single mode fiber), except for very large strain events, such as earthquakes, that are outside the scope of this paper (as they would be above the DAS saturation limit). Furthermore, the birefringence orientation $\theta_m$ does not change. \\On the contrary, the sum $\delta n_{x,m} + \delta n_{y,m} = -\frac{n_0^3}{2}((p_{11}+p_{12})(-2\nu)\epsilon_z+2p_{12}\varepsilon_z)$ is non-zero, so the common phase is affected both by a change in length and by photoelasticity. 
This modification of the common phase factor is what is typically detected in conventional DAS systems, for which the change in optical path length due to longitudinal strain is typically written: $\delta (\beta_{0,m}z) = \varepsilon_z\xi\beta_{0,m}z$ where $\xi \approx 0.78$ is the photoelastic scaling factor \cite{mecozzi_sensing_2024} that encompasses the changes both in refractive indices and length. Finally, while longitudinal strain only scales the existing retardance, it does not induce additional birefringence. However, as we will see in the next sub-section, this is not the case for events that are not axisymmetric, such as transversely applied forces or fiber bending, for which $\varepsilon_x \neq \varepsilon_y$.\\ 
\subsubsection{Transverse strain}
\label{sec:transverse strain}
For a wide range of perturbations, $\varepsilon_x \neq \varepsilon_y$, inducing birefringence. It would be the case, for instance, for a transverse load $f$ (in N/m) applied on the fiber, with an induced birefringence $\Delta\beta_{event} = \frac{2\pi}{\lambda}C\frac{4f}{\pi r E}$ \cite{huard_polarisation_1994} and a fast axis along the force direction, or for fiber bending, for which $\Delta\beta_{event} = \frac{2\pi}{\lambda}\frac{1}{2}C\frac{r^2}{R^2}$ \cite{yasin_optical_2012,ulrich_bending-induced_1980} where $C$ is computed from the photoelasticity theory, $r$ is the fiber radius and $E$ is the Young modulus of silica, with a fast axis normal to the bending plane. 

\begin{figure}[h]
    \centering
    \includegraphics[width=0.9\linewidth]{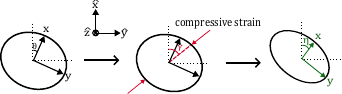}
    \caption{Example of a compressive transverse strain at an angle $\gamma$ from the reference frame $\hat{x}$ axis and resulting fiber deformation}
    \label{fig:transverse_strain}
\end{figure}

In this work, we focus on the differentiation of transverse and longitudinal strain. Hence, we apply an arbitrary anisotropic event, choosing $\varepsilon_x = -2\varepsilon_y$ and for simplicity $\varepsilon_z = 0$. This would be consistent for instance with a compressive strain, expanding the fiber in the orthogonal direction. This induces an additional birefringence  \begin{equation}
    \Delta \beta_{event} = -\frac{2\pi}{\lambda}\frac{n_0^3}{2}(p_{11} - p_{12})(\varepsilon_y - \varepsilon_x) 
\end{equation} for which the fast axis will be oriented along the compressive strain direction. The angle of the compressive strain with respect to the $\hat{x}$ axis of the reference frame is chosen arbitrarily, and the resulting birefringence parameters $\Delta \beta_i '$ and $\eta_i$ are computed as described in section~\ref{perturbations}. A representation of the considered deformation is represented in Fig. \ref{fig:transverse_strain}.
Given the expression of $\Delta \beta_i'$ in Equation~\eqref{resulting_biref}, the resulting magnitude will strongly depend on the angle between the orientation of the intrinsic birefringence and the orientation of the event-induced birefringence, leading to possible cancellation of birefringence. The resulting magnitude is not linear in the perturbation \cite{mecozzi_sensing_2024}, except when the two directions are aligned or orthogonal. However, its estimation can still provide information on the nature of the strains acting on the fiber. Additionally, we compute the change in common propagation constant $\beta_{0,i}$ due to transverse strain from Equation~\eqref{sum_n}. Indeed, while we don't model any change in physical length due to the considered transverse event, there is still a change in optical path length from the photoelastic effect.

It is worth noting that in most DAS systems, any change in common phase is interpreted as resulting from pure longitudinal strain, and this equivalent longitudinal strain $\varepsilon_z$ is computed using the formula:  $\delta (\beta_{0}z) = \varepsilon_z\xi\beta_{0}z$, which is actually not valid if the strain is not purely axial. According to the previously derived equations, the response in common phase due to transverse strain is actually quite different.

\section{DAS simulation model}
\label{sec:DAS_model}
\begin{figure}[h]
    \centering
    \includegraphics[width=1\linewidth]{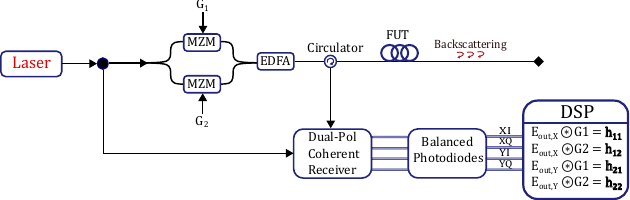}
    \caption{MIMO-DAS with coded-sequence based probing. MZM: Mach-Zehnder Modulator, EDFA: Erbium-Doped Fiber Amplifier, FUT: Fiber Under Test, DSP: Digital Signal Processing}
    \label{fig:DAS}
\end{figure}

We developed a MATLAB\textregistered ~model to emulate, as realistically as possible, the operation of the coded-DAS system displayed in Fig.~\ref{fig:DAS}. 

At the transmitter side, Polarization Division Multiplexed (PDM) BPSK-mapped Golay sequences with certain orthogonality properties\cite{dorize_enhancing_2018}, denoted $G_1$ and $G_2$, are used as probing signals to modulate a continuous-wave laser. They are then transmitted over a $3$km-SSMF fiber through an optical circulator.  
Each sequence contains $32768$ BPSK symbols (Golay order 12) emitted at a symbol rate of $200$~MBaud. Hence, the code length is $0.163$~ms, resulting in a mechanical bandwidth of $3.05$ kHz. We probe the fiber by sending $100$ codes successively in time. Using Golay sequences interrogation instead of a single pulse spreads the transmitted energy over time, significantly reducing the peak optical power of the probing signal, hence achieving the same sensing Signal-to-Noise Ratio (SNR) at equal transmitted average power. At the same time, this peak-power reduction enables co-existence of communication data and sensing signals by drastically reducing nonlinear interference between DAS and communication channels. An EDFA amplifier compensates the losses of the transmitter while adding ASE noise to the signal. The laser phase noise is modeled as a Wiener process of variance $\sigma^2 = 2\pi\Delta\nu T_s$, where the laser linewidth $\Delta\nu$ is set to $100$ Hz and the symbol period $T_s$ is $0.005$ $\mu$s. The signal power at the fiber input is $4.45$ dBm. The fiber is divided into $11757$ segments seen as waveplates of length $dz = 0.255$m each ($dz \ll L_C,\, L_b$).

To model a dynamic event, we generate, for each transmitted probing code $\mathbf{c}_l$ with $l:1\rightarrow100$, a list of round-trip matrices $ \mathbf{H}(l) = [\mathbf{H}_1^{rt}(l), \mathbf{H}_2^{rt}(l), ..., \mathbf{H}_{N_{seg}}^{rt}(l)]$ following the model described in section \ref{sec:fiber_model}. We insert an event at segment $i$ by computing in time (hence for each code), the birefringence parameter $\Delta\beta_i'$, $\eta_i$ and $\beta_{0,i}'$ of matrix $\mathbf{U}_i$ and of $\mathbf{U}_{i,scat}$, which results in modifying $\mathbf{H}_n^{rt}$ for $n = i$ to $n = N_{seg}$. 
The received signal is obtained by processing each transmitted code individually. For each code, the transmitted field $\mathbf{E}^{c_l}_{in}= \begin{pmatrix}
    E^{c_l}_{in, x} \\
    E^{c_l}_{in,y}
\end{pmatrix}$ is convolved with its corresponding impulse response $\mathbf{H}(l)$. The resulting signals from all codes are then added together in time, ensuring that overlapping contributions from different codes are correctly combined. Finally, only the portion of the summed signal corresponding to the original sequence length is retained, discarding any additional padding introduced by the convolution. This procedure provides an accurate representation of the full received sequence, taking into account the interference between successive codes.

The resulting backscattered signal $\mathbf{E}^{c_l}_{out}$ is detected at the receiver through a dual-polarization coherent detection. The signal level at the input of the coherent receiver is $-24$ dBm, and the local oscillator power is $4.47$ dBm. The receiver model also adds white Gaussian noise of variance $\sigma^2_{RX}~= ~1.23.10^{-5}\mathrm{A}^2$ representing thermal noise, shot noise and relative intensity noise (RIN) contributions from photodiodes and transimpedance amplifiers (TIA). To estimate the differential phase and the birefringence strength of each waveplate, several steps of post-processing are performed. First, the output of the back-propagation of each code is a mixture of the backscattering of every symbol of the code on every waveplate. To retrieve the unmixed output for each sequence, a correlation with the original code is performed~\cite{dorize_enhancing_2018}. This allows the estimation of the round-trip matrices list $\widehat{\mathbf{H}}(l) = \begin{pmatrix}
    \mathbf{h_{11}}(l) & \mathbf{h_{12}}(l) \\ \mathbf{h_{21}}(l) & \mathbf{h_{22}}(l)
\end{pmatrix}$ for each code $\mathbf{c}_l$ \cite{guerrier_introducing_2020}. 

\begin{figure*}[!b]
\centering
\subfloat{
\begin{overpic}[width=0.47\linewidth]{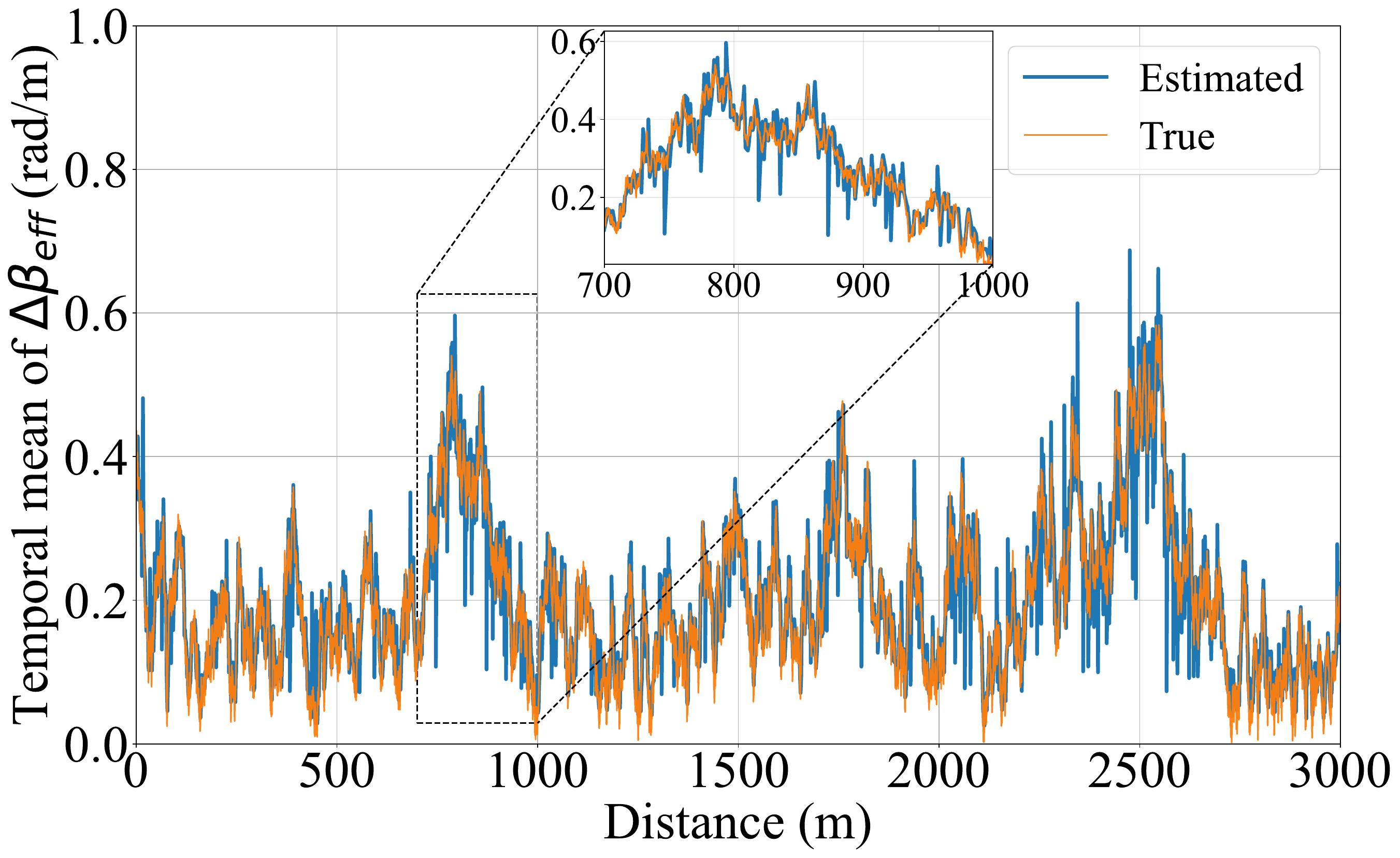}
    \put(-5,57){{(a)}}
\end{overpic}
\label{fig:biref_estimation}
}%
\hfil
\subfloat{
\begin{overpic}[width=0.44\linewidth]{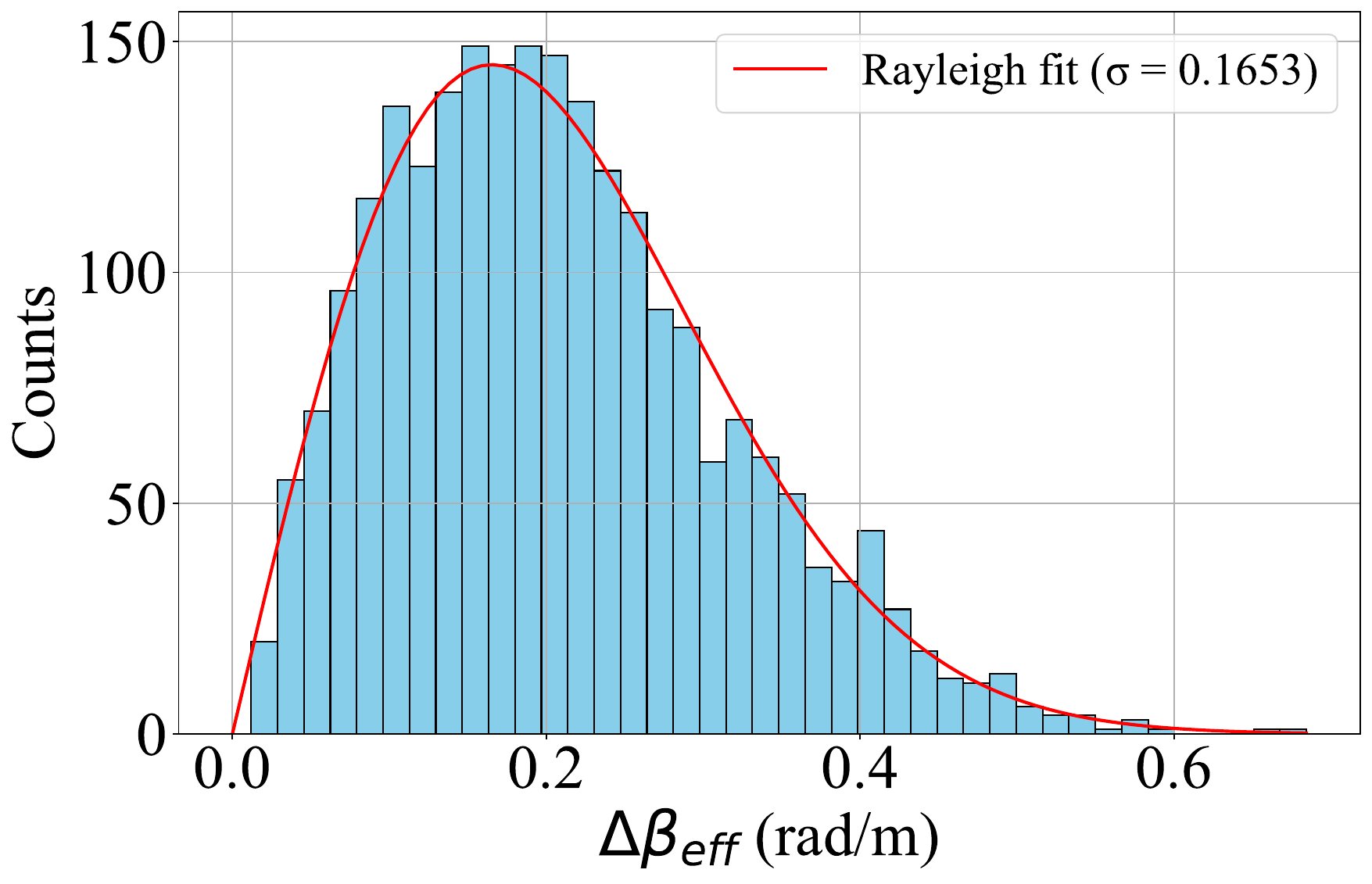}
    \put(-1,60){{(b)}}
\end{overpic}
\label{fig:biref_hist}
}
\caption{Estimation of the effective birefringence magnitude in static conditions: (a) Temporal mean of estimated effective birefringence magnitude along fiber length. (b) Histogram of estimated effective birefringence magnitude in the fiber and Rayleigh fit.}
\label{fig:biref_static}
\end{figure*}
To mitigate the error introduced by coherent fading that leads to some segments contributions to the backscattered wave having very low values (which will induce large estimation errors), we select, after the correlation at the receiver side, the highest reflecting segments \cite{guerrier_high_2022} every $5$ consecutive segments, downsampling the number of studied segments to $N_{red}=2351$, and obtaining a mean spatial resolution or mean gauge length of $1.28$ m. We index the selected matrices with $n = [1 ,...,N_{red}]$. This non-uniform downsampling step selects segments that are not necessarily separated by the same length (it can vary from $0.255$m to $2.55$m for a low-resolution factor of $5$), which is why the resolution is described as a mean resolution. It results in a non-uniform gauge length, and the estimated quantities described in the following are therefore effective quantities on this non-uniform gauge length. For small amplitude events, to get a better contrast between the event and ambient noise, higher low-resolution factors can be used, for instance selecting the highest reflective segments amongst groups of 10. The choice of the low-resolution factor is a trade-off between localization accuracy and sensitivity.

From each code $\mathbf{c}_l$, the common phase is estimated by taking 
\begin{equation}
    \phi_n(l) = 0.5\angle\det(\mathbf{H}_n^{rt}(l))\end{equation} as in \cite{guerrier_introducing_2020}, for each selected matrix $n$.  In the following, we drop the index $l$ to simplify the notations. To obtain the differential phase $\Delta \phi_n$, which is the quantity of interest for strain measurements, we take the spatial difference between the phase computed from each selected matrix and the previous one (separated by one gauge length).  

To retrieve the effective birefringence strength over the gauge length, we compute, as proposed in previous works~\cite{costa_localization_2023, galtarossa_reflectometric_2008, yaman_polarization_2023, feng_distributed_2018}, the product matrix \begin{equation}
    \mathbf{T}_n = (\widehat{\mathbf{H}}_{n-1}^{rt})^{\dagger}\widehat{\mathbf{H}}_{n}^{rt}
\end{equation} 
The two eigenvalues $\Lambda_{1,n}$ 
and $\Lambda_{2,n}$
of the product matrix are computed. Then, we compute $|0.5\angle(\Lambda_{1,n}\Lambda_{2,n}^*)|$ to remove common phase terms and only keep the effective retardance $\Delta\beta_{eff,n}L_n$ (the cumulated polarization effect). By dividing by the length of the fiber between the two selected segments $L_n$, we can obtain the effective birefringence magnitude $\Delta\beta_{eff,n}$ of the section, in rad/m. We use the term effective to emphasize the fact that it is actually the effective polarization rotation around the effective birefringence vector of the whole section~\cite{costa_localization_2023} and not the real local birefringence magnitude (which varies at a smaller length scale, $25$~cm in the simulation model).  These estimations of $\phi_n$ and $\Delta \beta_{eff,n}$ are performed for all segments and for all codes, providing a dynamic distributed estimation. 

\section{Simulation results}

In this section, we validate our model in three different scenarios: static conditions, dynamic longitudinal strain and dynamic transverse strain. We demonstrate its ability to localize, detect and discriminate events through the estimation of common phase and effective birefringence strength. 
\subsection{Static conditions}
\begin{figure*}[!t]
\centering
\subfloat{
    \begin{overpic}[width=0.45\linewidth]{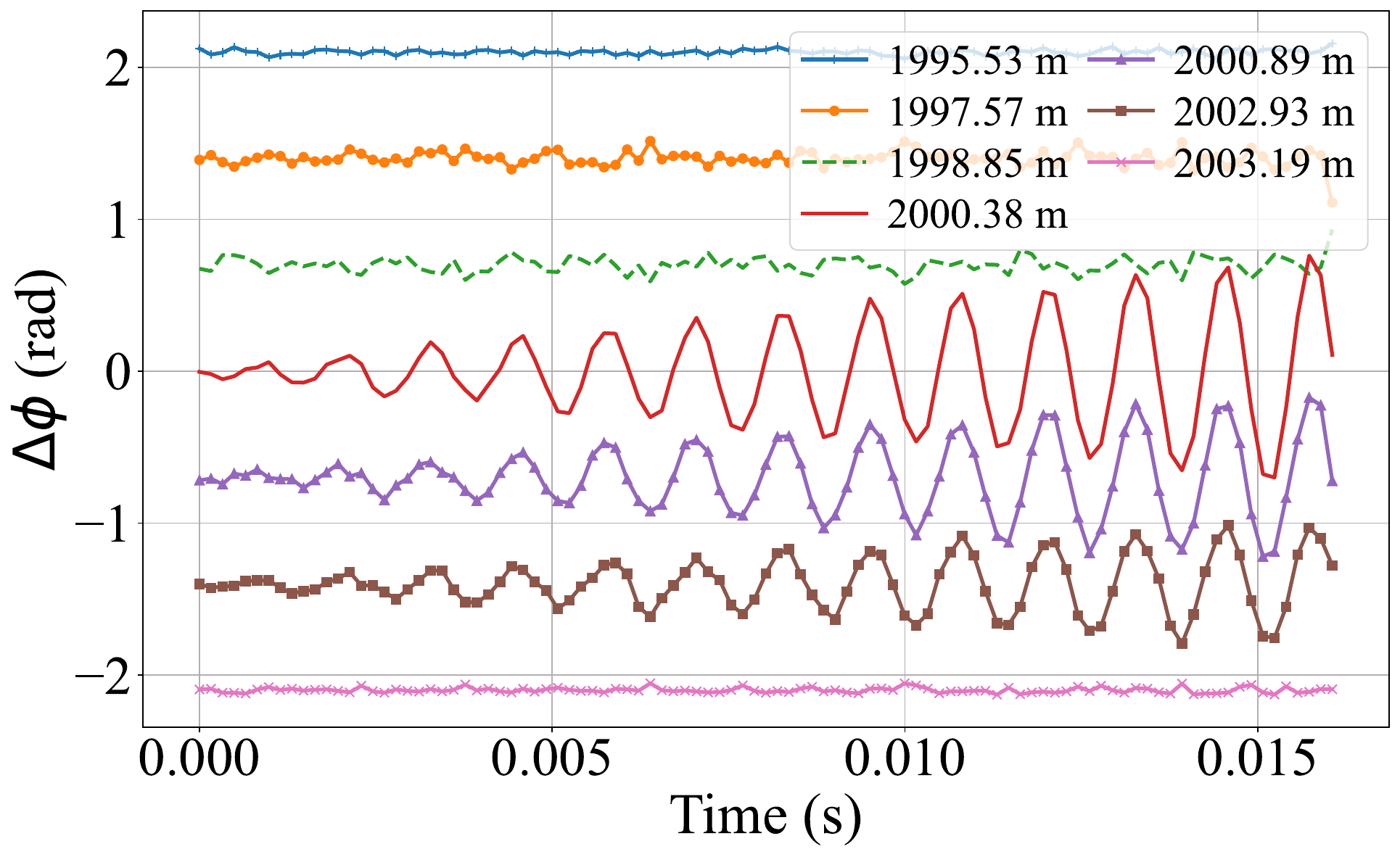}
        \put(0.1,57){{(a)}}
    \end{overpic}
    \label{fig:diff_phi_longitudinal}
}%
\hfil
\subfloat{
    \begin{overpic}[width=0.45\linewidth]{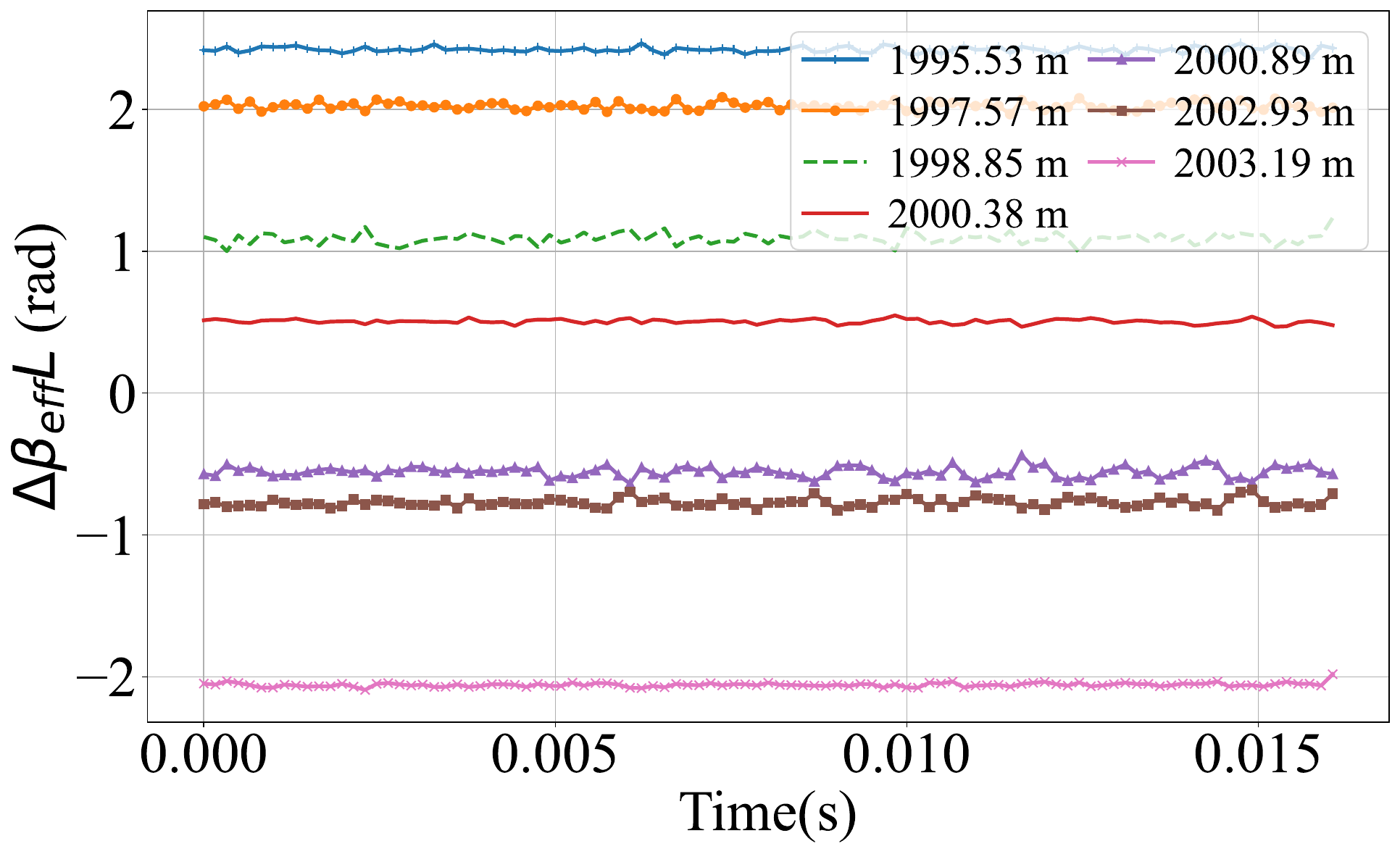}
        \put(0.1, 57){{(b)}}
    \end{overpic}
    \label{fig:betaL_longitudinal}
}
\caption{Estimated parameters under longitudinal strain at event position and neighboring segments: 
(a) Estimated differential phase $\Delta \phi$ (rad). 
(b) Estimated retardance $\Delta \beta_{eff}L$ (rad).}
\label{fig:disturbed_segment}
\end{figure*}
In static conditions, the distribution of $\Delta \beta_{eff}$ along the fiber length can be retrieved after dividing each $\Delta \beta_{eff,n}L_n$ by the proper distance $L_n$. The distance between two selected segments can vary between $0.255$m and $2.55$m since we choose the best segment amongst groups of 5. Fig.~\ref{fig:biref_estimation} shows the estimated effective birefringence magnitude over the fiber length (averaged over all the 100 codes). We observe that the estimated birefringence profile is close to the true applied values, up to some discrepancies. The discrepancies are mainly due to the fact that we are estimating over a gauge length of $1.3$m on average even though the local birefringence orientation is varying across this length (every $0.25$m in this simulation model), the estimated birefringence magnitude is an effective magnitude, and not the sum of the magnitudes of the local birefringence vectors. Indeed, the higher the low-resolution factor (the bigger the gauge length), the smaller the estimated mean birefringence value. Moreover, the eigenvalue method does not give an exact estimate since all the scatterers are not placed at the same position at the end of each segment, but distributed randomly in each segment. This leads to the appearance of a coherent sum in Equation~\eqref{Hroundtrip} that makes the birefringence estimation process from the product matrix $\mathbf{T}_n$ only an approximation. 
\begin{figure*}[!b]
\centering
\subfloat{
\begin{overpic}[width=0.45\linewidth]{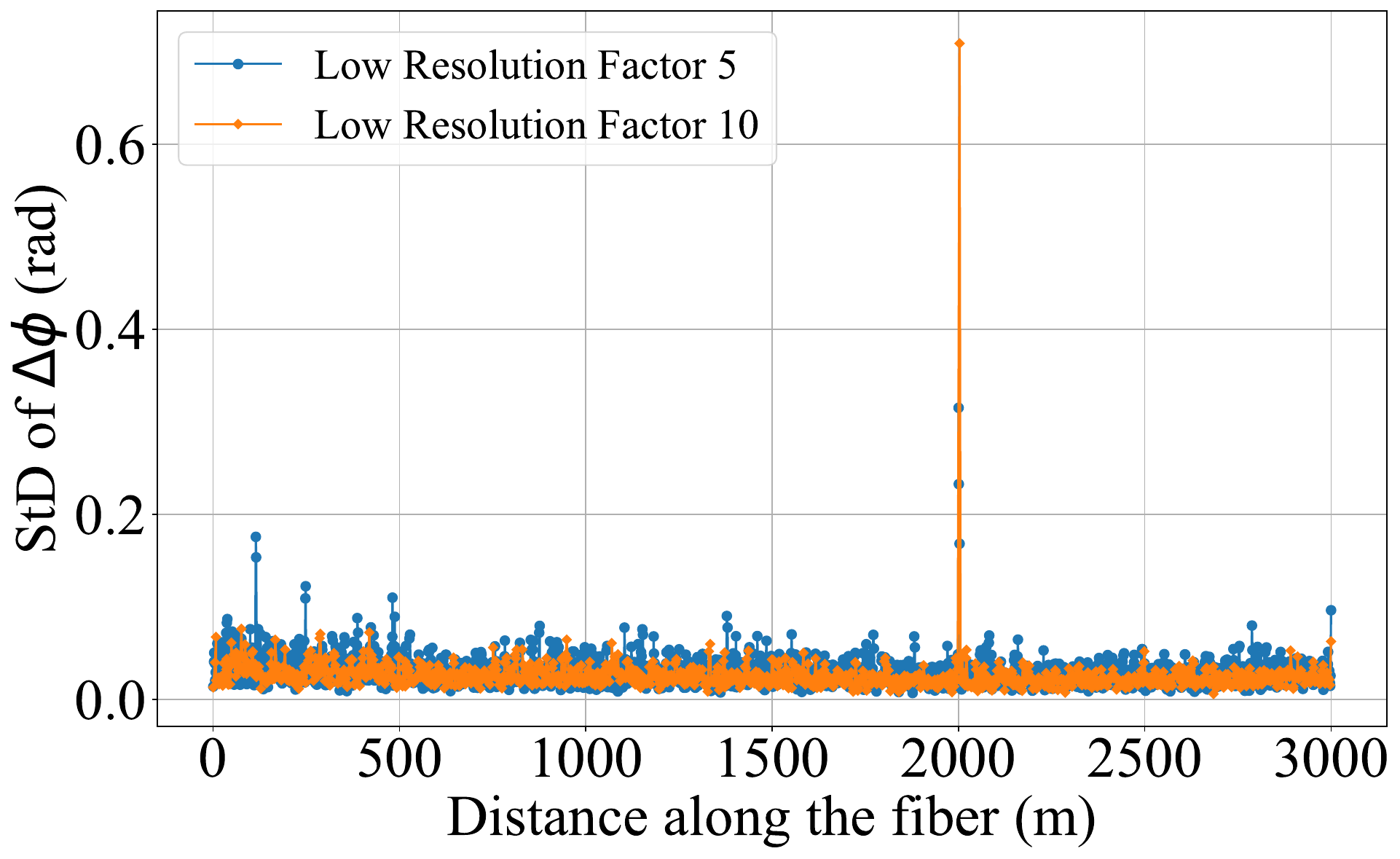}
    \put(0.1,57){{(a)}}
\end{overpic}
\label{fig:std_phi_longitudinal}
}
\hfil
\subfloat{
\begin{overpic}[width=0.45\linewidth]{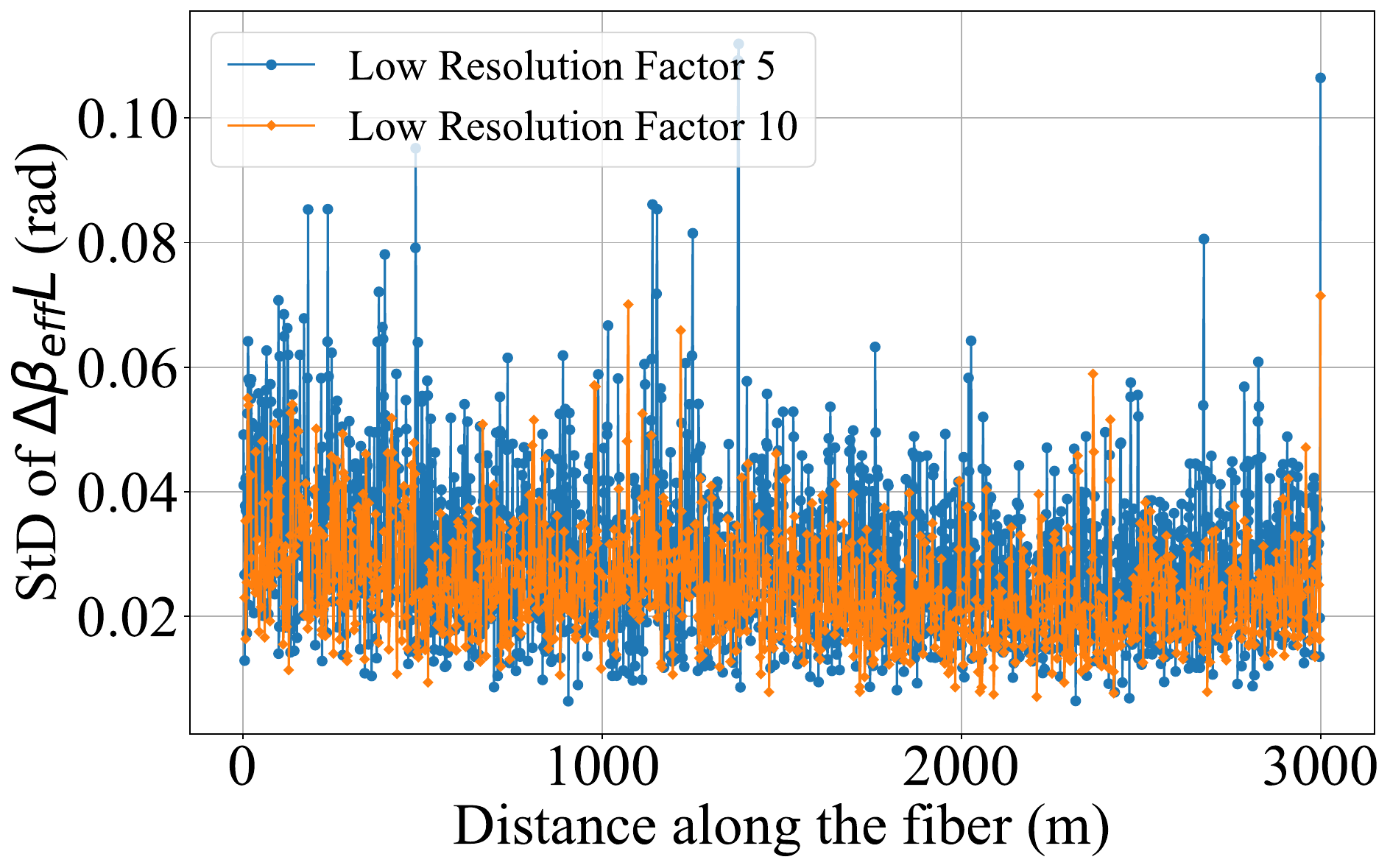}
    \put(0.1,57){{(b)}}
\end{overpic}
\label{fid:std_betaL_longitudinal}
}
\caption{Standard deviation of estimated parameters in the presence of longitudinal strain at $\approx 2000$ m: 
(a) Standard deviation of differential phase $\Delta \phi$ (rad). 
(b) Standard deviation of retardance $\Delta \beta_{eff}L$ (rad).}
\label{fig:std_longitudinal}
\end{figure*}

Fig. \ref{fig:biref_hist} represents the distribution of the birefringence magnitude along the fiber. We obtain a distribution for the estimated birefringence strength close to the expected Rayleigh distribution. The obtained mean value is $\approx 0.21$ rad/m, which is in agreement with the expected value \cite{wuilpart_measurement_2001}.

\subsection{Dynamic longitudinal strain}
To assess the effect of dynamic longitudinal strain, we apply an $800$ Hz sinusoidal wave (with an offset to apply only positive strain values) of varying amplitude from $10$ n$\varepsilon$ to $250$ n$\varepsilon$ on 5 fiber segments, hence spreading over $1.3$m in a $3$ km fiber, from $1999.9$m to $2001.2$ m. The model parameters are updated as explained in section \ref{longitudinal_strain}. 
Fig. \ref{fig:diff_phi_longitudinal} shows the estimated differential phase as a function of time at the impacted segment and neighboring segments (an offset is added to each trace for better visibility). We observe that the sine wave perturbation can be retrieved from $2000.38$ m  to $2002.93$ m and is not visible on the neighboring segments, confirming the ability of the differential phase estimation to isolate the event~\cite{guerrier_high_2022}. Conversely, as expected for a pure longitudinal strain, the event cannot be retrieved on the estimated retardance $\Delta \beta_{eff}L$ traces represented in Fig. \ref{fig:betaL_longitudinal}. 

Fig. \ref{fig:std_phi_longitudinal} and \ref{fid:std_betaL_longitudinal} show the standard deviations of the differential phase $\Delta \phi$ and of the retardance $\Delta \beta_{eff} L$ for two different low-resolution factors: $5$ and $10$ segments. While the peak corresponding to the impacted segment is clearly observable on the differential phase standard deviation for both down-sampling settings, the event goes unnoticed on the standard deviation of $\Delta \beta_{eff}L$. Note that the number of peaks and their amplitude differ for the two low-resolution factors because of the selection process that will include more or less segments in the strained region. These results validate the model confirming that the common phase is sensitive to pure longitudinal strain, whereas the birefringence magnitude remains unaffected.

\subsection{Dynamic transverse strain}
To model the impact of a dynamic anisotropic strain, we apply, at the same location as previously considered, a sine wave on $\varepsilon_y$ varying from $2$ n$\varepsilon$ to $200$ n$\varepsilon$ (with an offset to keep it positive) with frequency $800$ Hz, and $\varepsilon_x = -2\varepsilon_y$ (as it would be the case for instance for a transverse loading for which a compression in x direction would result in an extension in the perpendicular direction) \cite{tu_theoretical_2025}. The direction of the applied strain $\gamma$ is arbitrarily set to $20$ deg. The corresponding changes in birefringence and phase are computed according to sections \ref{photoelasticity} and \ref{perturbations}. 
\begin{figure*}[!t]
\centering
\subfloat{
\begin{overpic}[width=0.45\linewidth]{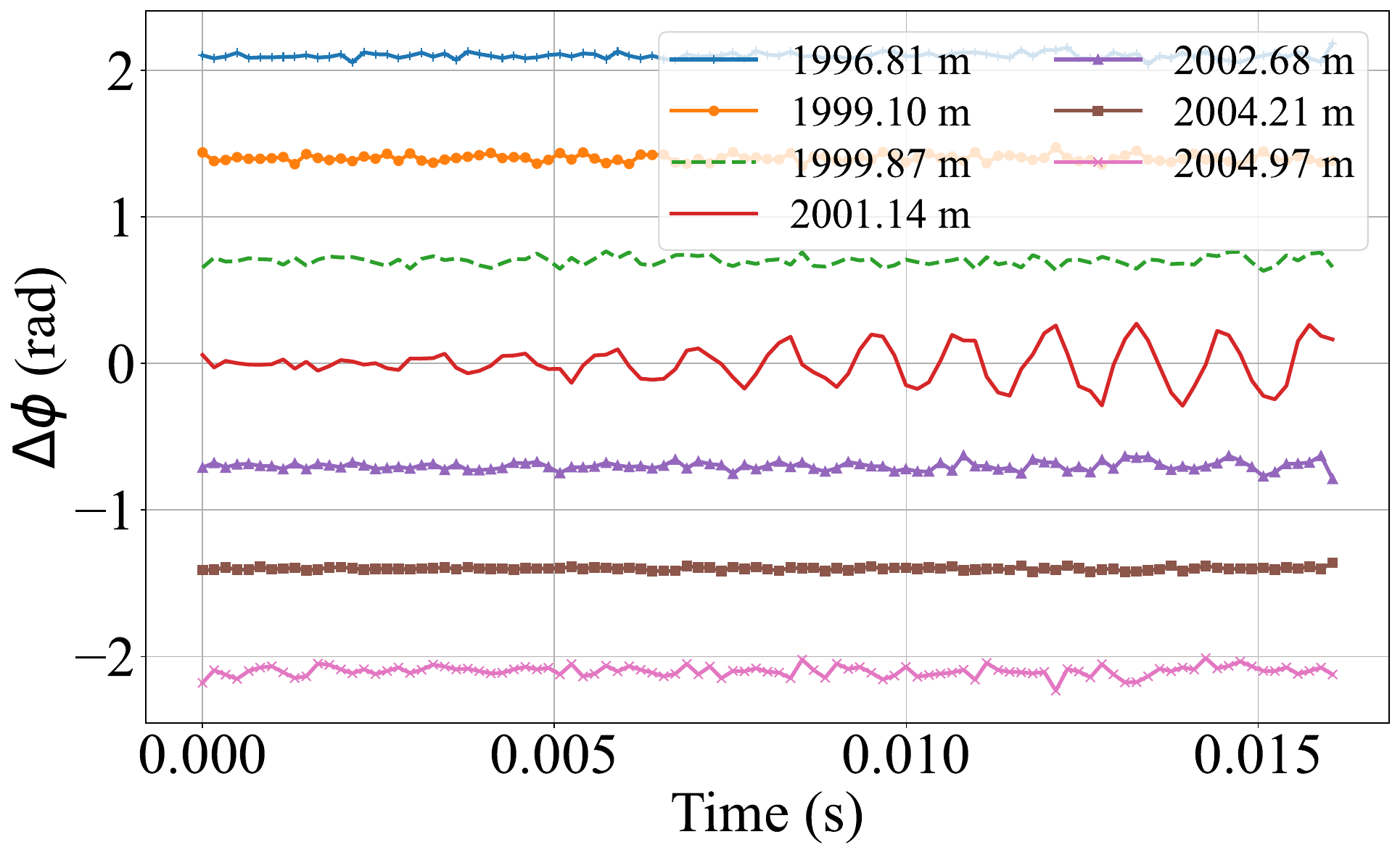}
    \put(0.1,57){{(a)}}
\end{overpic}
\label{fig:transverse_event_diffPhi}
}
\hfil
\subfloat{
\begin{overpic}[width=0.425\linewidth]{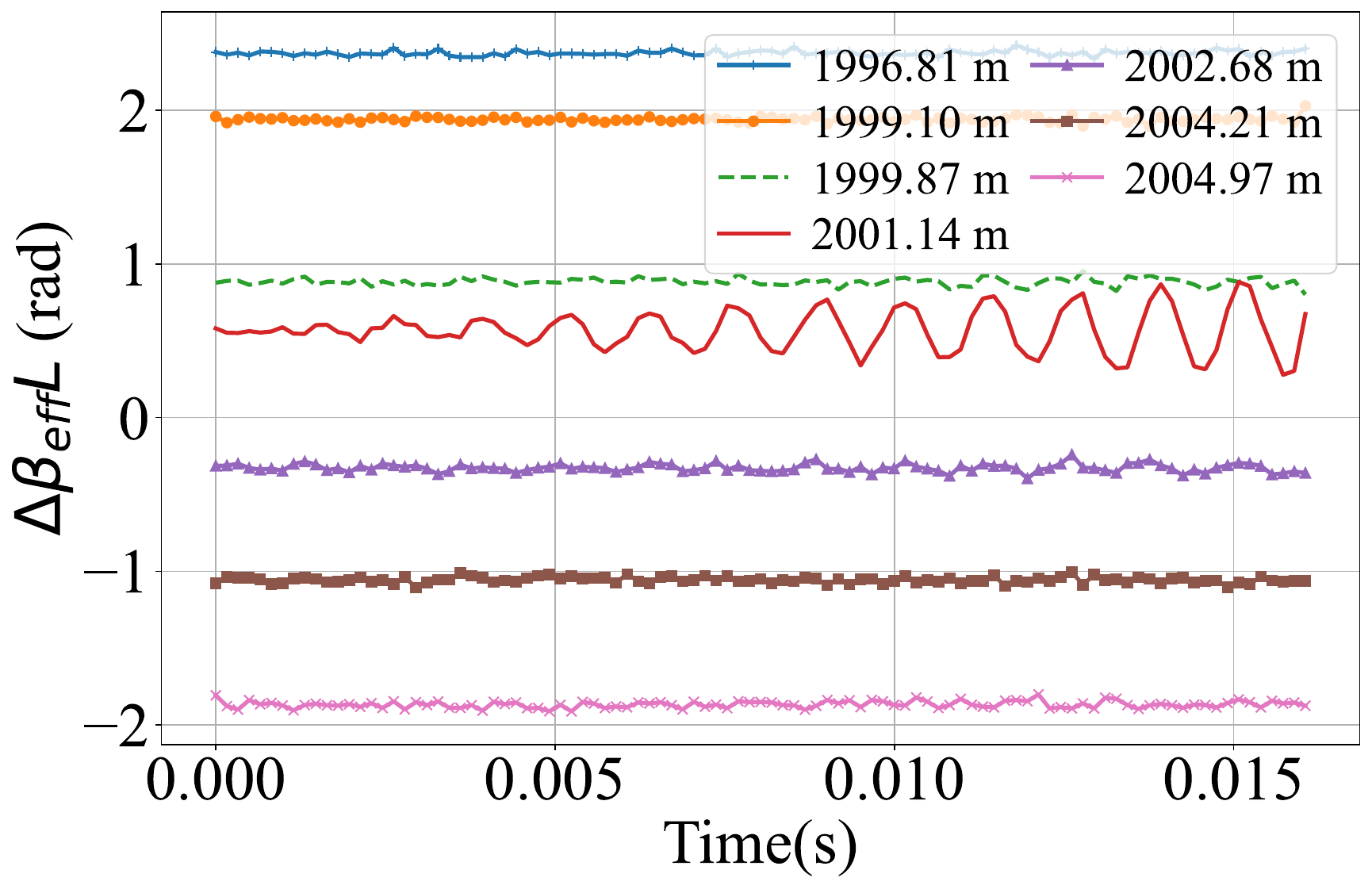}
    \put(0.1,60){{(b)}}
\end{overpic}
\label{fig:transverse_event_betaL}
}
\caption{Estimated parameters under transverse perturbation at event position and neighboring segments: 
(a) Estimated differential phase $\Delta \phi$ (rad). 
(b) Estimated retardance $\Delta \beta_{eff}L$ (rad).}
\label{fig:transverse_event}
\end{figure*}
\begin{figure*}[b!]
\centering
\subfloat{
\begin{overpic}[width=0.45\linewidth]{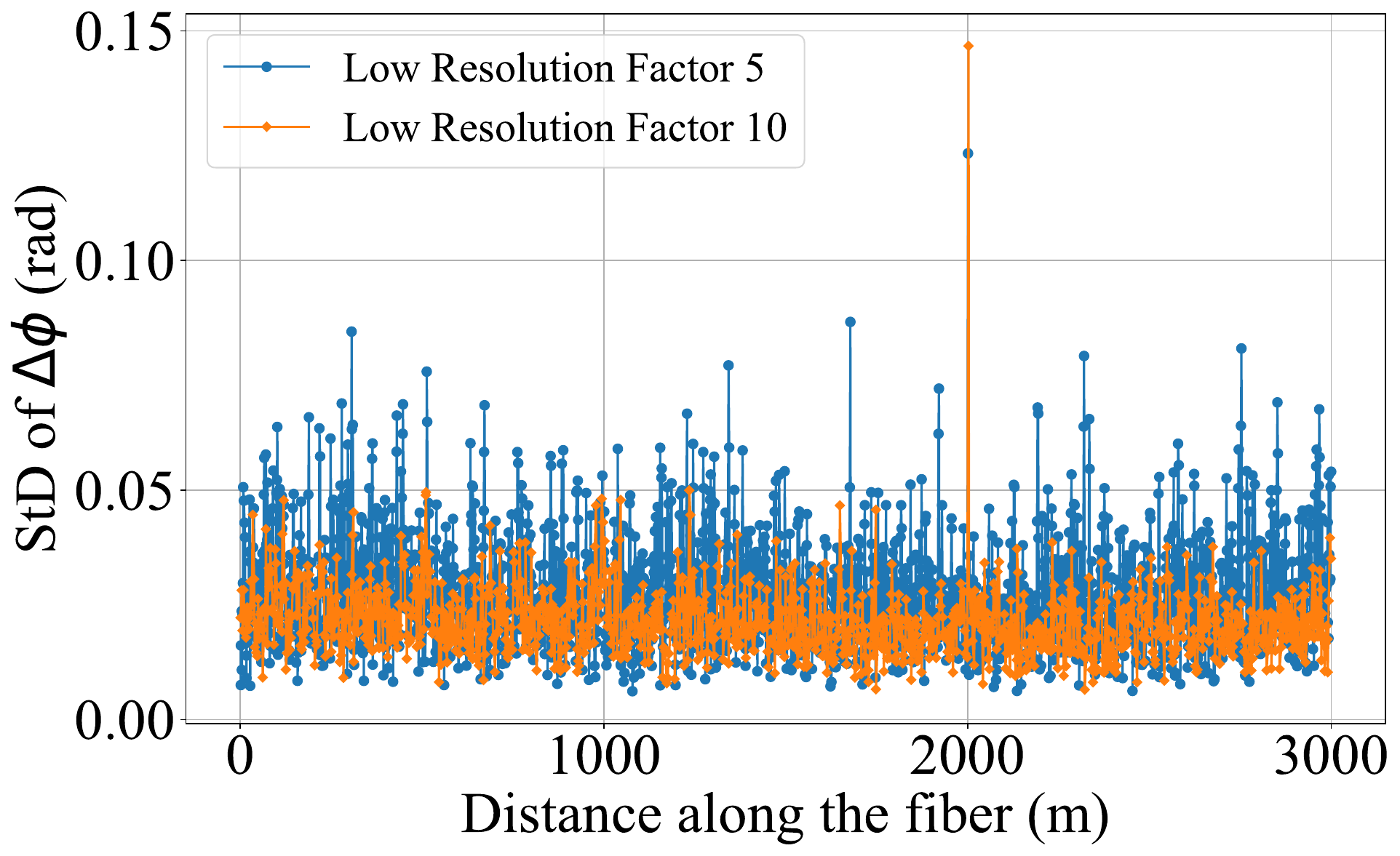}
    \put(0.05,57){{(a)}}
\end{overpic}
\label{fig:std_diffPhi_transverse}
}
\hfil
\subfloat{
\begin{overpic}[width=0.45\linewidth]{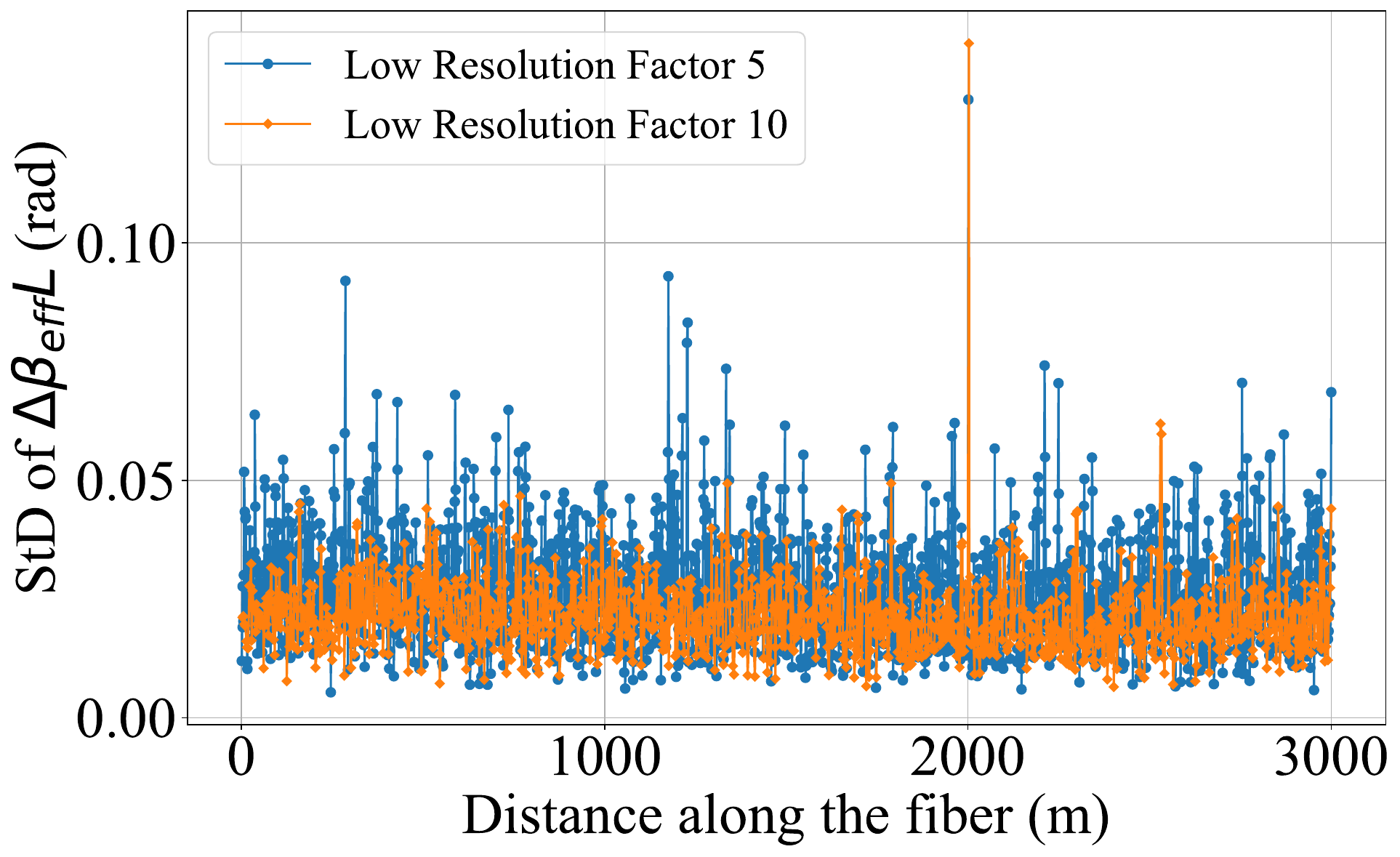}
    \put(0.1,57){{(b)}}
\end{overpic}
\label{fig:std_betaL_transverse}
}
\caption{Standard deviation of estimated parameters in the presence of transverse strain at $\approx 2000$ m: 
(a) Standard deviation of differential phase $\Delta \phi$ (rad). 
(b) Standard deviation of retardance $\Delta \beta_{eff}L$ (rad).}
\label{fig:std_transverse}
\end{figure*}

Fig. \ref{fig:transverse_event_diffPhi} and \ref{fig:transverse_event_betaL} demonstrate that the anisotropic perturbation is localized at position $2001.14$ m through both phase and birefringence magnitude estimations, as opposed to longitudinal strain which was only detectable on the common phase trace.  Fig. \ref{fig:std_diffPhi_transverse} and \ref{fig:std_betaL_transverse} show that the standard deviations of both differential phase and retardance exhibit peaks corresponding to the event position, for low-resolution factors of $5$ (mean gauge length $1.3$m) and $10$ (mean gauge length $2.6$m). A better contrast is achieved when selecting highest reflecting segments among groups of ten. Hence, the choice of the low-resolution factor defines a trade-off between sensitivity and spatial resolution and can be adjusted depending on the targeted performance.

Overall, the developed and validated model shows that given the nature of the applied dynamic strain, the behavior of the differential phase and effective birefringence magnitude recovered from a MIMO-DAS system differs: while pure longitudinal strain only impacts the differential phase estimate, anisotropic transverse strain affects both parameters. This observation paves the way for discrimination between longitudinal and transverse strains applied on a fiber.  

\section{Experimental results}
Experimental measurements were performed to validate the model in lab environment. We explore the system behavior both in static and dynamic conditions.

\subsection{Static measurements}

\begin{figure*}[t!]
\centering
\subfloat{
\begin{overpic}[width=0.45\linewidth]{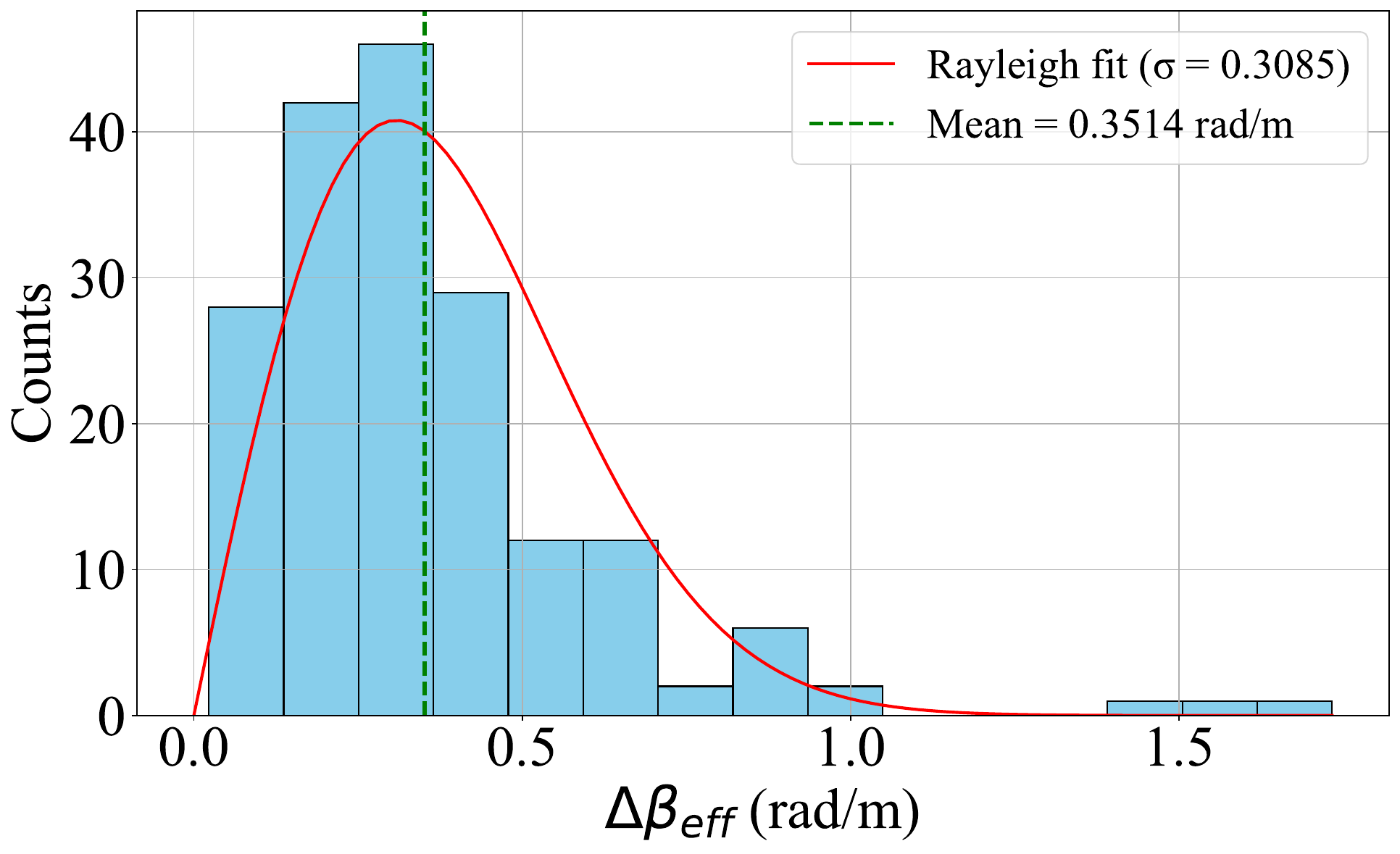}
    \put(0.05,57){{(a)}}
\end{overpic}
\label{fig:histogram_235m}
}
\hfil
\subfloat{
\begin{overpic}[width=0.45\linewidth]{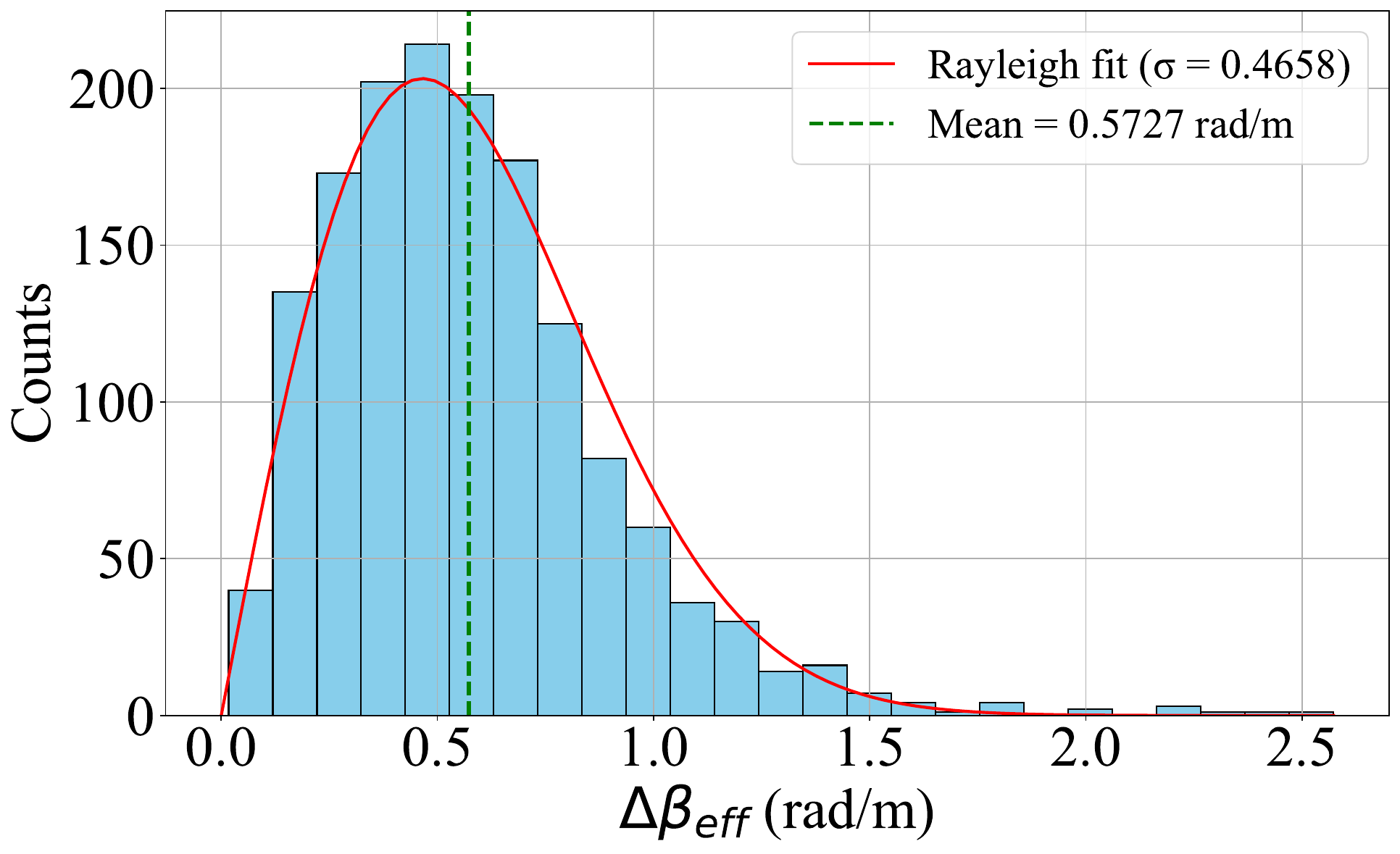}
    \put(0.1,57){{(b)}}
\end{overpic}
\label{fig:histogram_1958m}
}
\caption{Histogram of estimated effective birefringence magnitude in the fiber and Rayleigh fit with a mean spatial resolution of $1.3$m:
(a) 235m fiber spool.
(b) 1958m fiber spool.}
\label{fig:histogram_spools}
\end{figure*}

First, we estimated the effective birefringence magnitude distribution along fiber length in two different fiber spools, of $235$m and $1956$m respectively. Similarly to the simulation conditions, we send Golay codes of duration $0.163$~ms (Golay order 12) and the low resolution factor is set to $5$, leading to a mean spatial resolution of approximately $1.3$m. Figure \ref{fig:histogram_spools} shows the distribution of the effective birefringence magnitude for each fiber spool estimated from one of the probing codes, fitted with a Rayleigh distribution. We observe that the distribution for the $235$m fiber represented in Figure \ref{fig:histogram_235m} is not close to a Rayleigh distribution, which can be explained by its insufficient length. Conversely, the histogram for the $1956$m fiber spool reasonably fits a Rayleigh distribution of scale parameter $\sigma \approx 0.47$. The two fibers have different mean values of effective birefringence magnitude, namely $\approx 0.35$rad/m for the $235$m fiber and $\approx 0.57$rad/m, which could be explained by different factors such as tension around the spool, pressure introduced by fiber layers on the spool or different internal stresses involuntarily introduced during fabrication... It is worth noting that the obtained values depend on the spatial resolution as observed in~\cite{chen_distributed_2021}. Indeed, a small spatial resolution would result in a strong impact of coherent fading with a tendency of increasing the estimated values, while larger spatial steps would cause a decrease in the effective birefringence magnitude due to an averaging effect with the change of birefringence orientation.

\begin{figure}[h]
    \centering
    \includegraphics[width=1\linewidth]{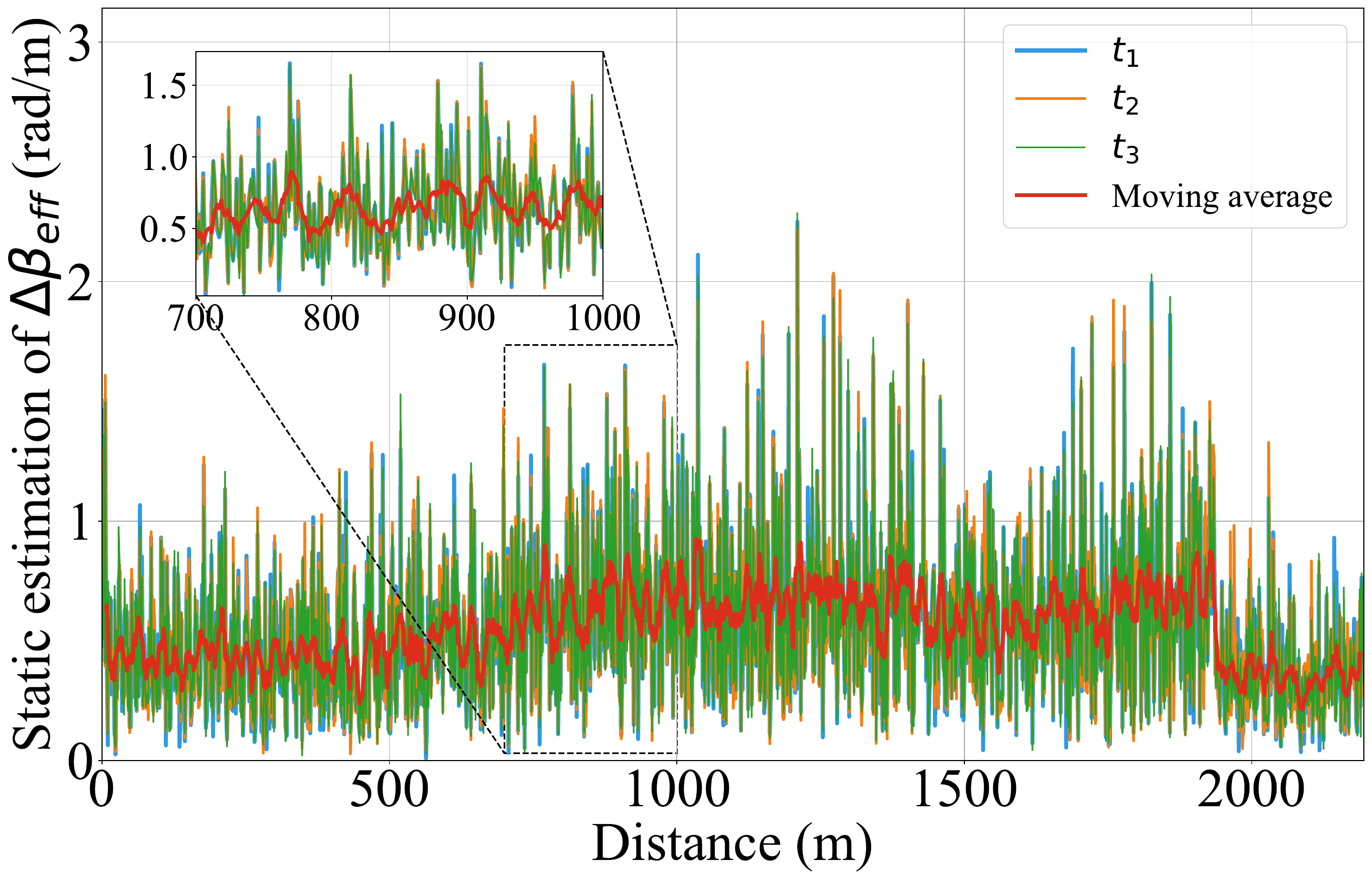}
    \caption{Estimation of the effective birefringence magnitude in static conditions for 3 different time instants and moving average with window size 15. The transition between the two spools is clearly visible around $1960$~m.}
    \label{fig:birefringence_vs_distance_exp}
\end{figure}

While the absolute values of the estimates depend on the system parameters, their statistical properties can be of interest to differentiate between different fibers or between fiber cable environments. Figure \ref{fig:birefringence_vs_distance_exp} shows the static effective birefringence strength profile obtained at three different time instants $t_1$, $t_2$ and $t_3$ when concatenating the two fiber spools ($1956$m followed by $235$m). A spatial moving average of window size $15$ is plotted to highlight the main trends. We can see clearly the transition from one fiber to the other, and the mean values computed from each spool agree with the ones measured on the fiber spools independently, demonstrating the ability to differentiate between them.

\subsection{Dynamic measurements}
\begin{figure}[h]
    \centering
    \includegraphics[width=0.9\linewidth]{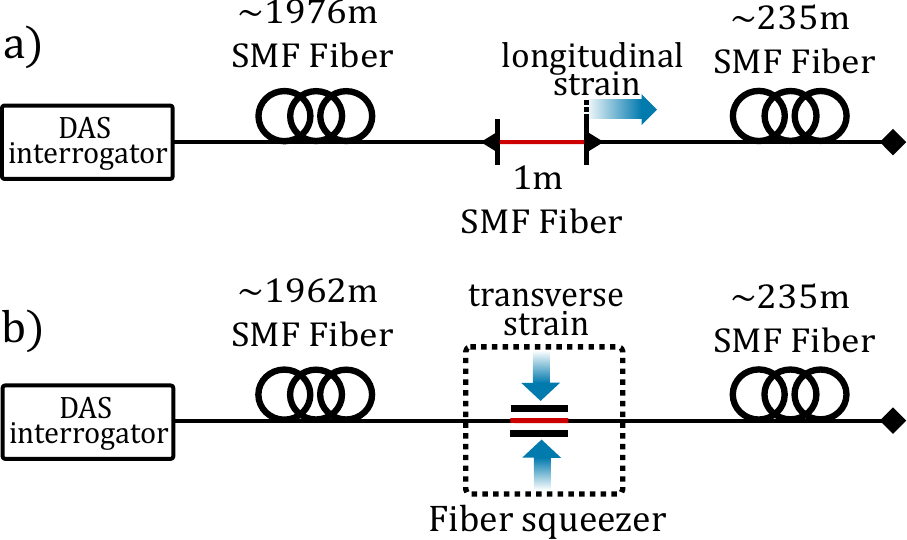}
    \caption{Experimental setup: (a) longitudinal strain is applied on a 1m fiber section. The 1m fiber is pulled from one end. (b) transverse strain is applied by using a fiber squeezer that squeezes the fiber transversely along an axis.}
    \label{fig:exp_setup}
\end{figure}

To assess the system ability to discriminate between longitudinal and transverse strains, we applied both events on fiber sections as shown in Figure \ref{fig:exp_setup}. To produce a longitudinal strain event as pure as possible, we stretched a one meter fiber horizontally at $1977$m along the longitudinal fiber axis and fixed its ends. Its right end is pulled by a moving stage controlled by a piezoelectric device to which we apply a positive sinusoidal tension. After the strained section, we connect the $235$m fiber spool, as per Figure \ref{fig:exp_setup}(a). To apply transverse strain, we use a commercial fiber squeezer that introduces linear birefringence by squeezing a small region of fiber (sub-centimeter scale) transversely. It is placed at approximately $1962$m as per Figure \ref{fig:exp_setup}(b). We apply a sinusoidal control voltage. Piezoelectric device constraints limit the event frequency to $100$Hz, but our DAS system is theoretically able to detect events up to $3.05$kHz. Moreover, the used fiber squeezer exhibits an unusual behavior even at low amplitude (hysteresis and nonlinear response confirmed through a reference measurement using a polarimeter), thus limiting the strength of the applied disturbance. 
\begin{figure*}[!t]
\centering
\subfloat{
\begin{overpic}[width=0.45\linewidth]{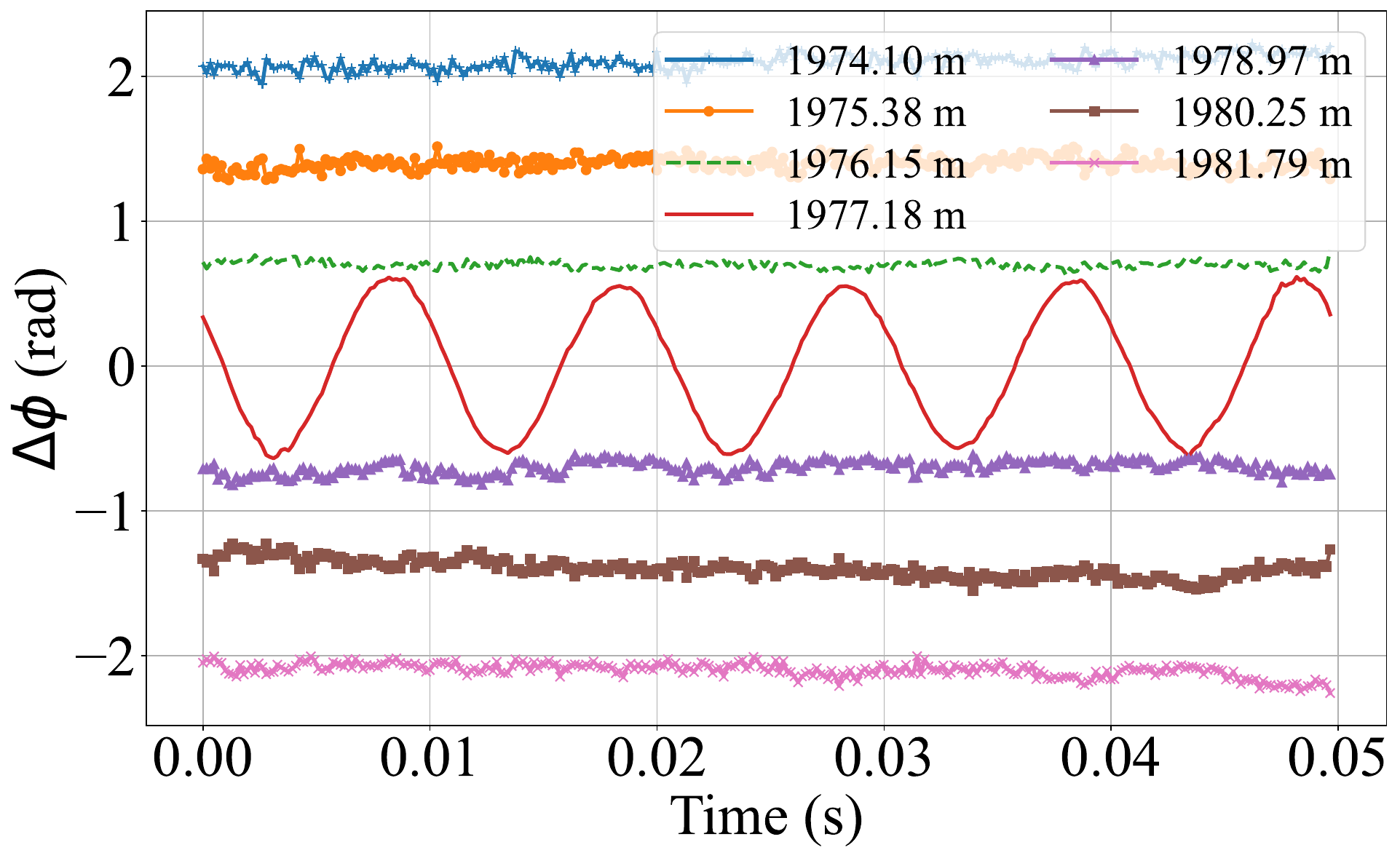}
    \put(0.1,57){{(a)}}
\end{overpic}
\label{fig:longitudinal_event_diffPhi_exp}
}
\hfil
\subfloat{
\begin{overpic}[width=0.425\linewidth]{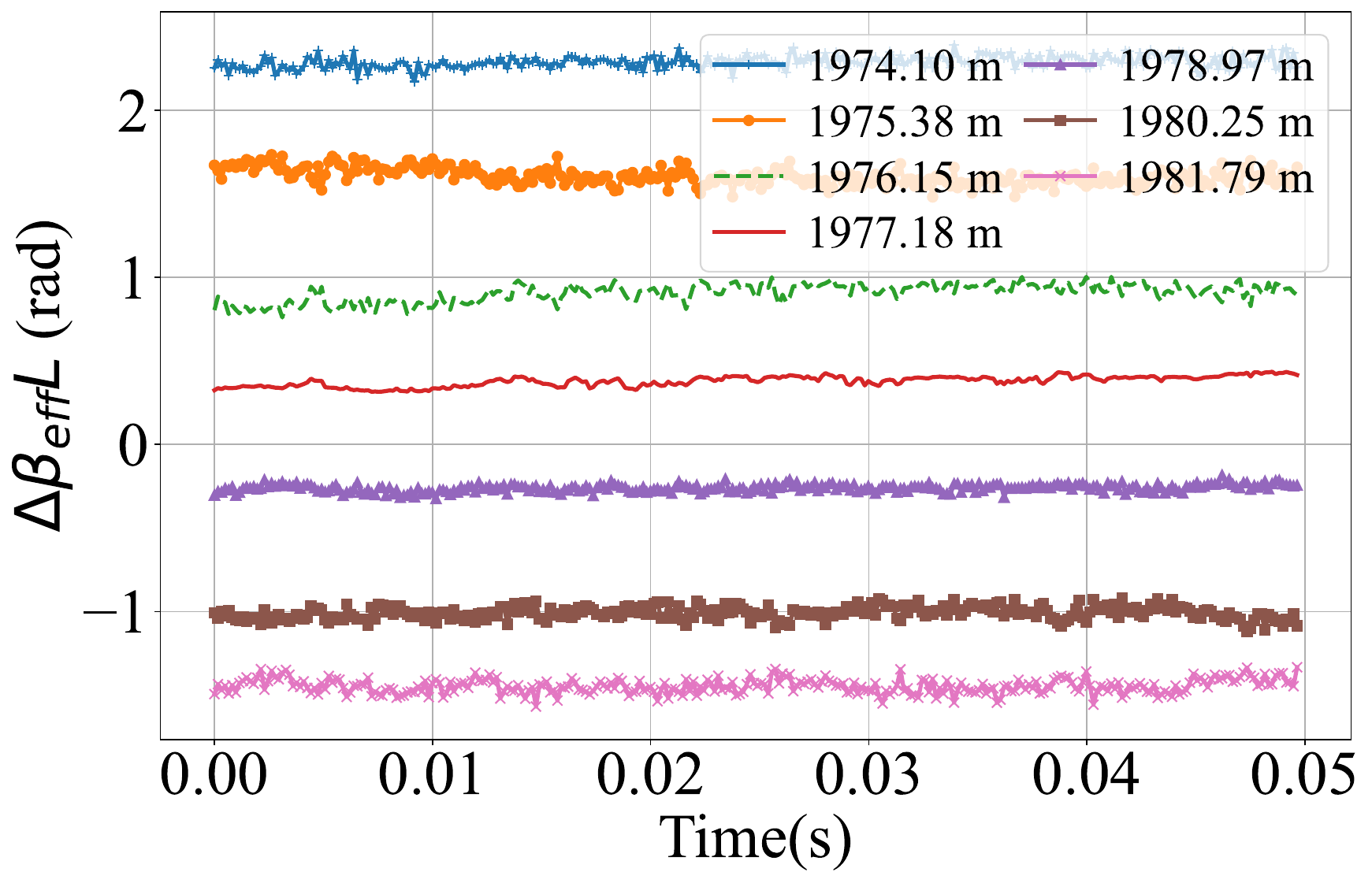}
    \put(0.1,60){{(b)}}
\end{overpic}
\label{fig:longitudinal_event_betaL_exp}
}
\caption{Estimated differential phase $\Delta \phi$ in (a) and retardance $\Delta \beta_{eff}L$ in (b) for a longitudinal perturbation at event position and at neighboring segments.}
\label{fig:longitudinal_event_exp}
\end{figure*}
\begin{figure*}[!t]
\centering
\subfloat{
\begin{overpic}[width=0.43\linewidth]{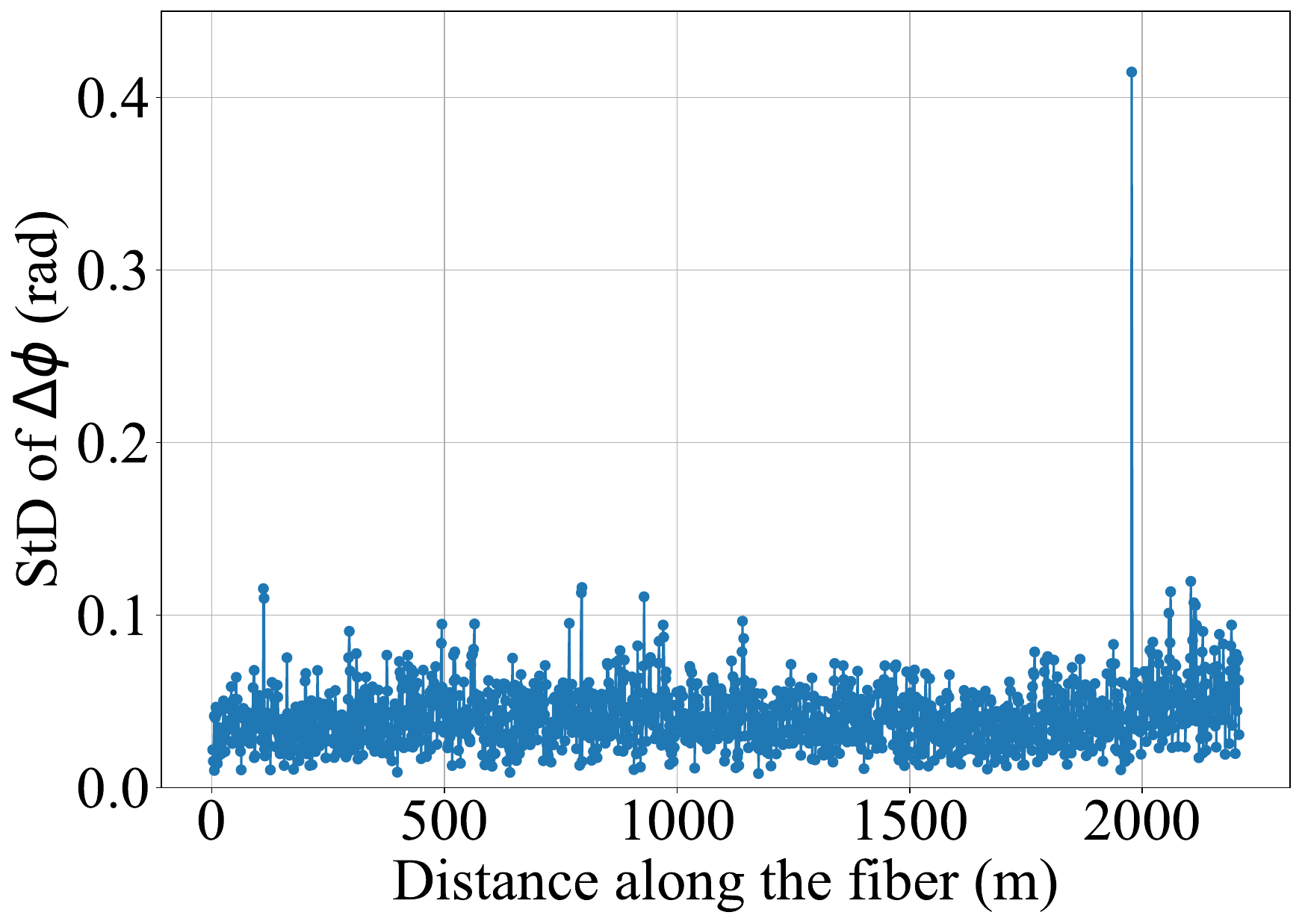}
    \put(0.1,67){{(a)}}
\end{overpic}
\label{fig:Std_longitudinal_diffPhi_exp}
}
\hfil
\subfloat{
\begin{overpic}[width=0.455\linewidth]{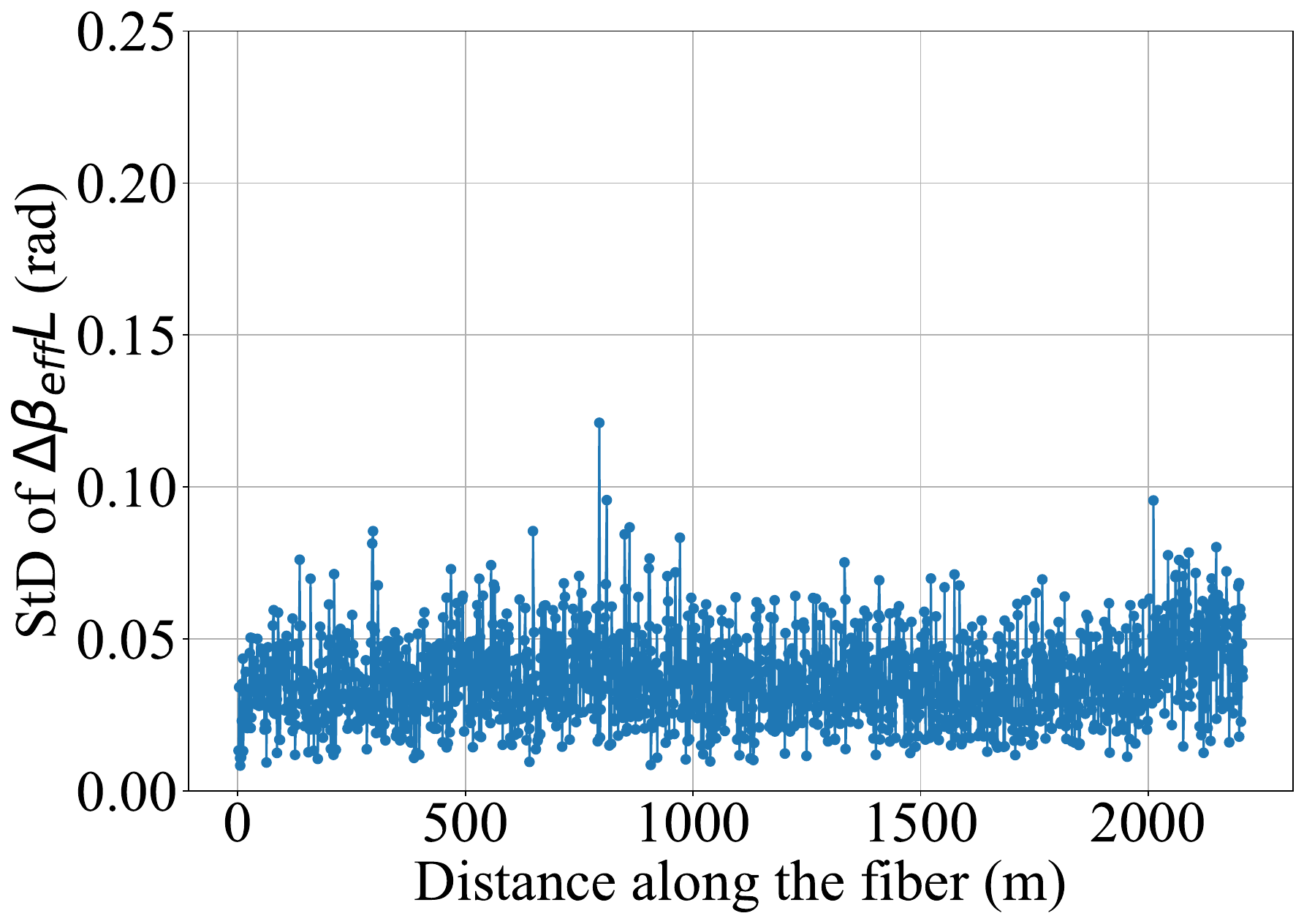}
    \put(0.1,63){{(b)}}
\end{overpic}
\label{fig:Std_longitudinal_betaL_exp}
}
\caption{Standard deviation (StD) of estimated parameters in presence of longitudinal strain: 
(a) StD of differential phase $\Delta \phi$, 
(b) StD of retardance $\Delta \beta_{eff}L$.}
\label{fig:Std_longitudinal_event_exp}
\end{figure*}

\subsubsection{Longitudinal strain} 
Longitudinal strain is applied by pulling one meter of fiber from one end in the previously described setup. A $100$Hz sine wave strain is introduced. Figure \ref{fig:longitudinal_event_exp} displays traces of the estimated differential phase $\Delta \phi$ and effective retardance $\Delta\beta_{eff}  L$ in time at different positions around the event position. The sinusoidal signature of the strain event is clearly observed at $1977.18$m on the differential phase estimation in Figure \ref{fig:longitudinal_event_diffPhi_exp}, while it is not noticeable on the retardance traces in Figure \ref{fig:longitudinal_event_betaL_exp}, demonstrating the strong effect of longitudinal strain on the differential phase parameter and the insensitivity of the retardance to this strain event. The standard deviation of the differential phase in Figure \ref{fig:Std_longitudinal_diffPhi_exp} exhibits a peak at the event position, proving its ability to locate the event with no impact on previous or next positions. 

\subsubsection{Transverse strain}
\begin{figure*}[!t]
\centering
\subfloat{
\begin{overpic}[width=0.45\linewidth]{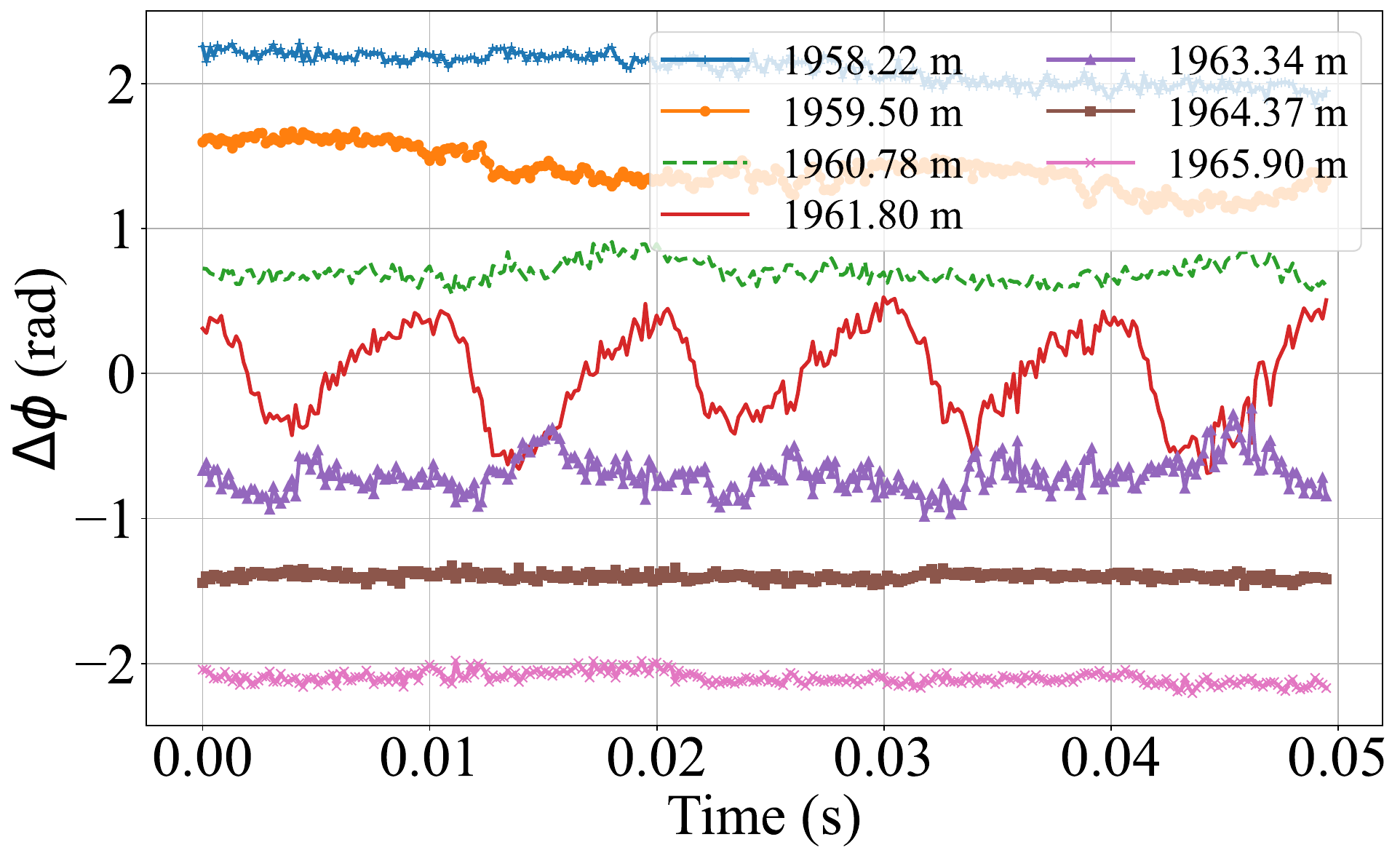}
    \put(0.1,57){{(a)}}
\end{overpic}
\label{fig:transverse_event_diffPhi_exp}
}
\hfil
\subfloat{
\begin{overpic}[width=0.43\linewidth]{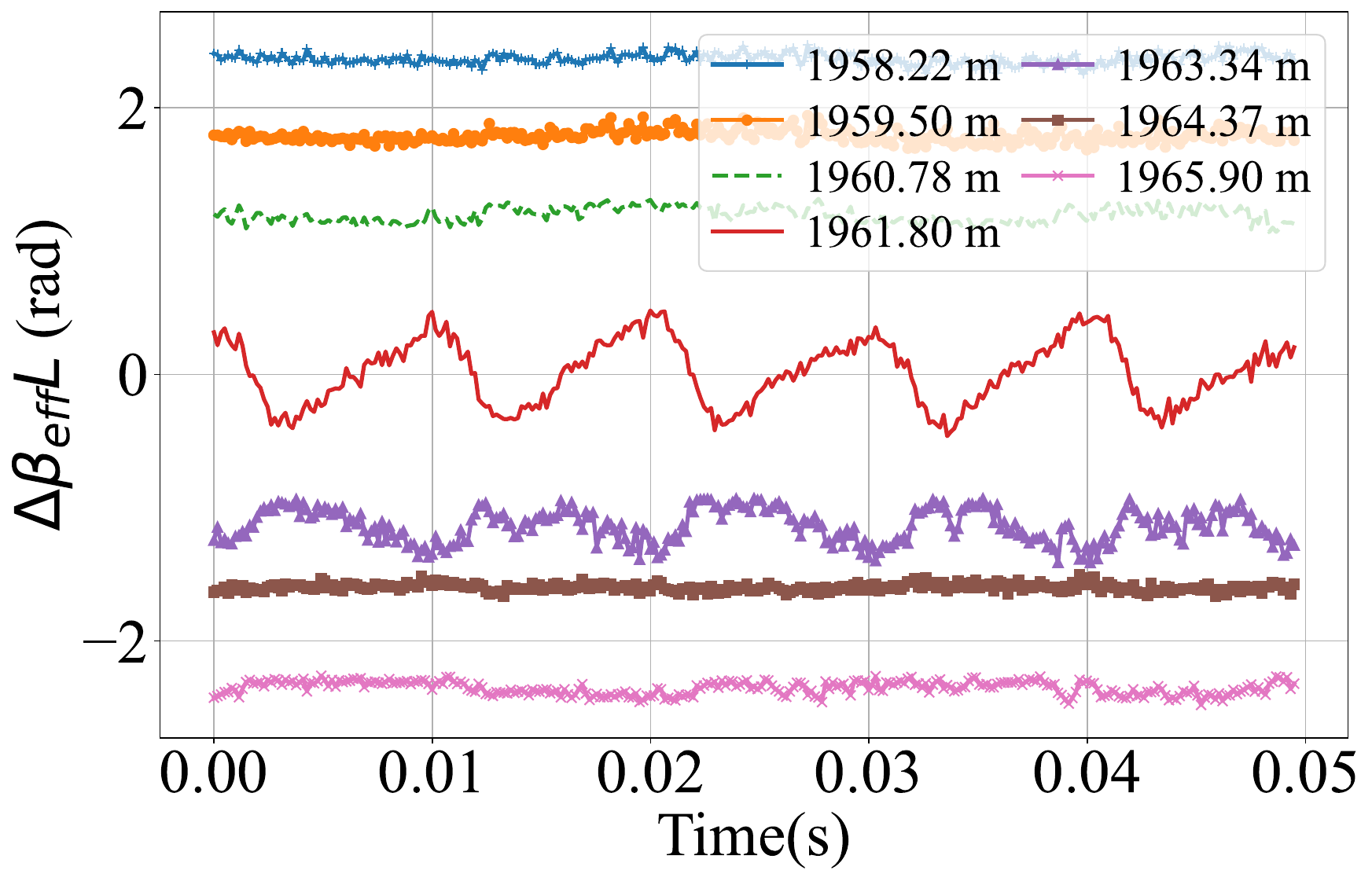}
    \put(0.1,60){{(b)}}
\end{overpic}
\label{fig:transverse_event_betaL_exp}
}
\caption{Estimated differential phase $\Delta \phi$ in (a) and retardance $\Delta \beta_{eff}L$ in (b) for a transverse perturbation at event position and at neighboring segments.}
\label{fig:transverse_event_exp}
\end{figure*}

\begin{figure*}[!t]
\centering
\subfloat{
\begin{overpic}[width=0.435\linewidth]{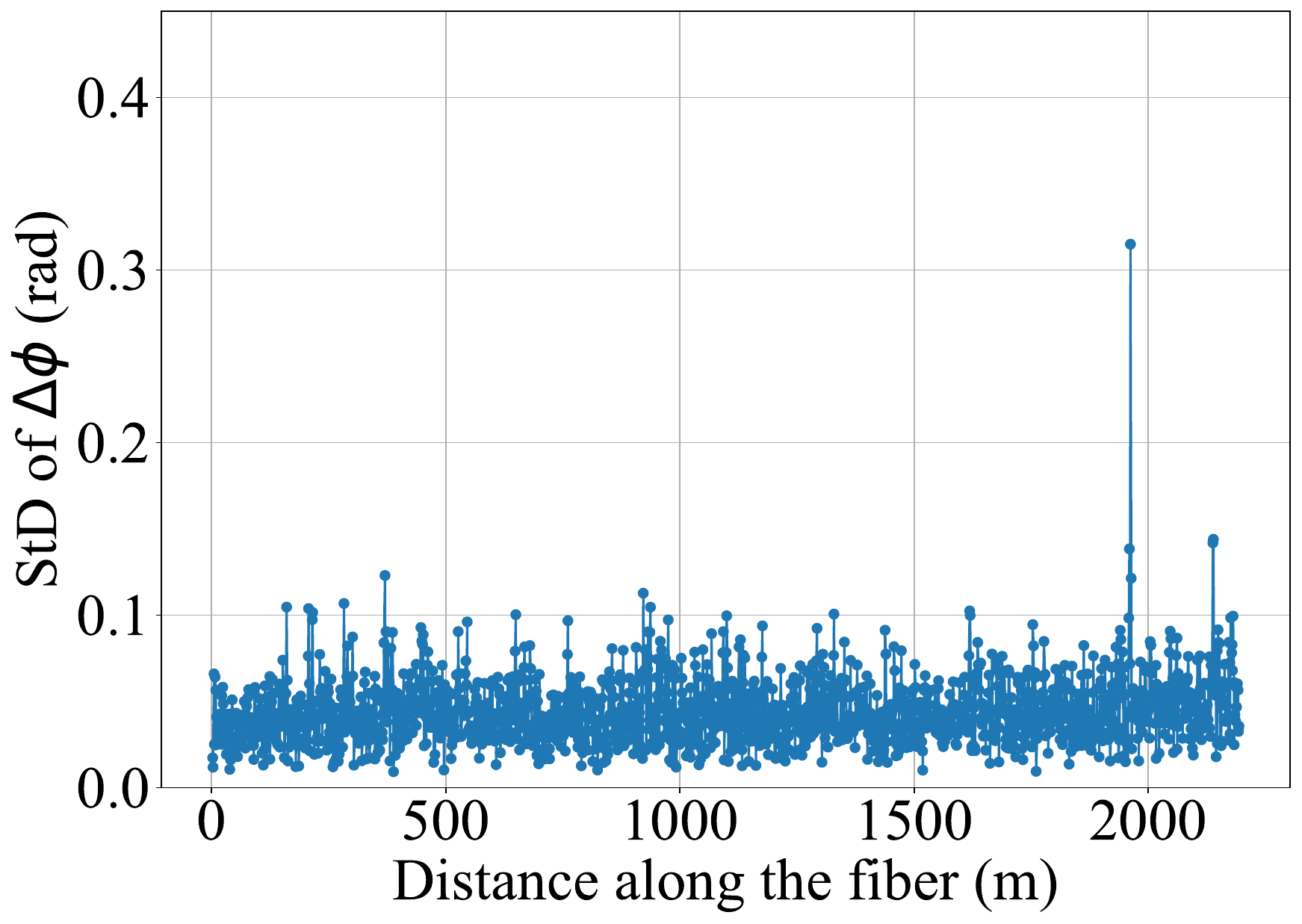}
    \put(0.1,67){{(a)}}
\end{overpic}
\label{fig:Std_transverse_diffPhi_exp}
}
\hfil
\subfloat{
\begin{overpic}[width=0.45\linewidth]{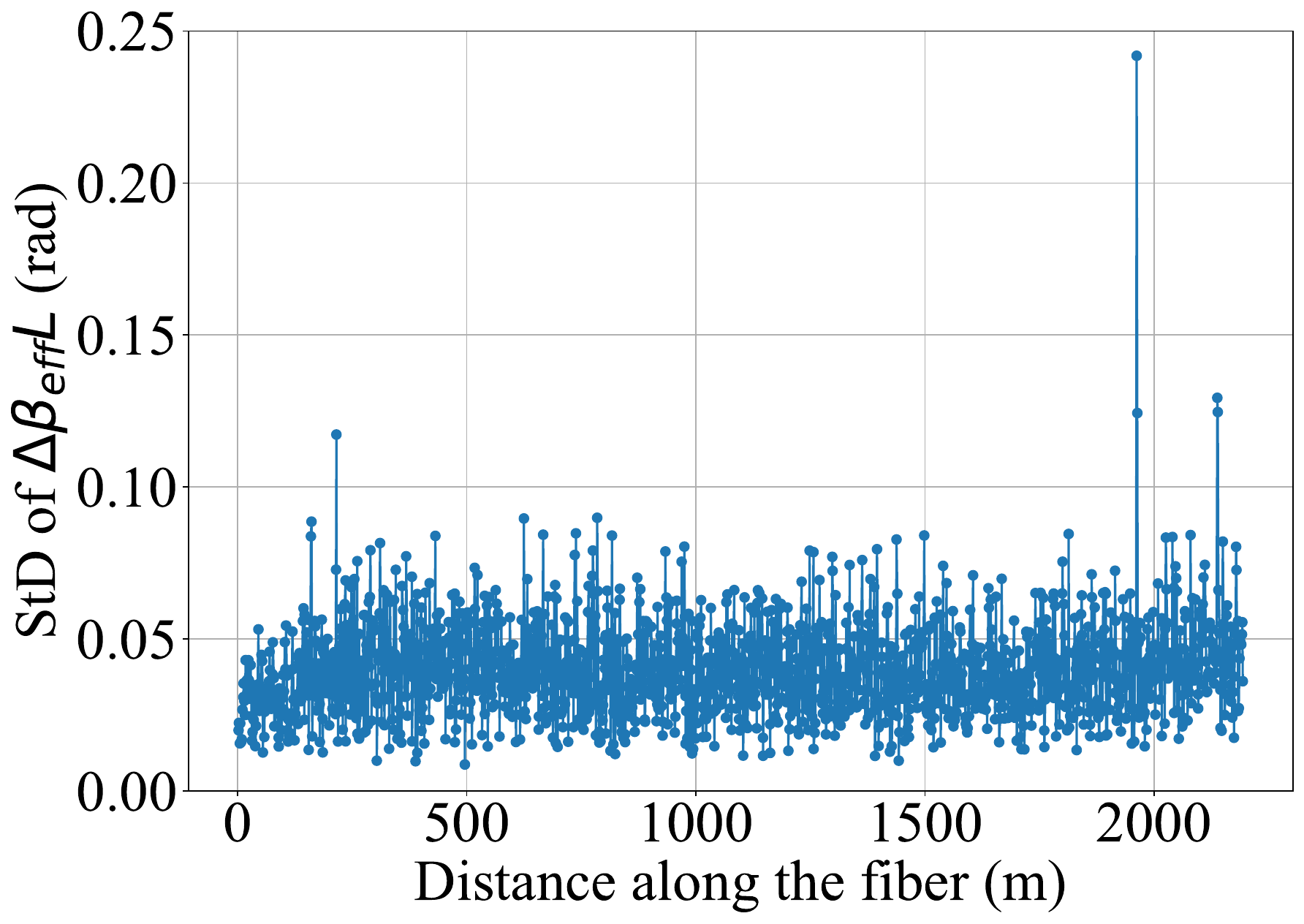}
    \put(0.1,65){{(b)}}
\end{overpic}
\label{fig:Std_transverse_betaL_exp}
}
\caption{Standard deviation (StD) of estimated parameters in presence of transverse strain: 
(a) StD of differential phase $\Delta \phi$, (b) StD of retardance $\Delta \beta_{eff}L$}
\label{fig:Std_transverse_event_exp}
\end{figure*}

A fiber squeezer device is used to apply transverse force on the fiber: a small length of fiber (on the order of magnitude of a few mm) is squeezed between two parallel plates to introduce linear birefringence. We control the device with a $100$Hz sinusoidal voltage. The evolution of $\Delta \phi$ and $\Delta\beta_{eff}L$ with time at the event position and neighboring positions is represented on Figure \ref{fig:transverse_event_exp}. Periodic fluctuations at the event frequency are retrieved on both differential phase (Figure \ref{fig:transverse_event_diffPhi_exp}) and retardance (Figure \ref{fig:transverse_event_betaL_exp}) traces, in agreement with simulation results. The event signature is not a perfect sinewave but presents some hysteresis which is due to the used fiber squeezer (this hysteresis and signal distortion was also observed using a polarimeter in forward transmission). Nevertheless, the event is detected by both parameters at $1961.80$m, with a mean resolution of $1.3$m. The standard deviations of differential phase and retardance, shown in Figures \ref{fig:Std_transverse_diffPhi_exp} and \ref{fig:Std_transverse_betaL_exp} respectively, further demonstrate the localization ability of the system, with no impact on the remaining fiber length.

\section{Discussion}

\begin{figure}[h]
    \centering
    \includegraphics[width=1\linewidth]{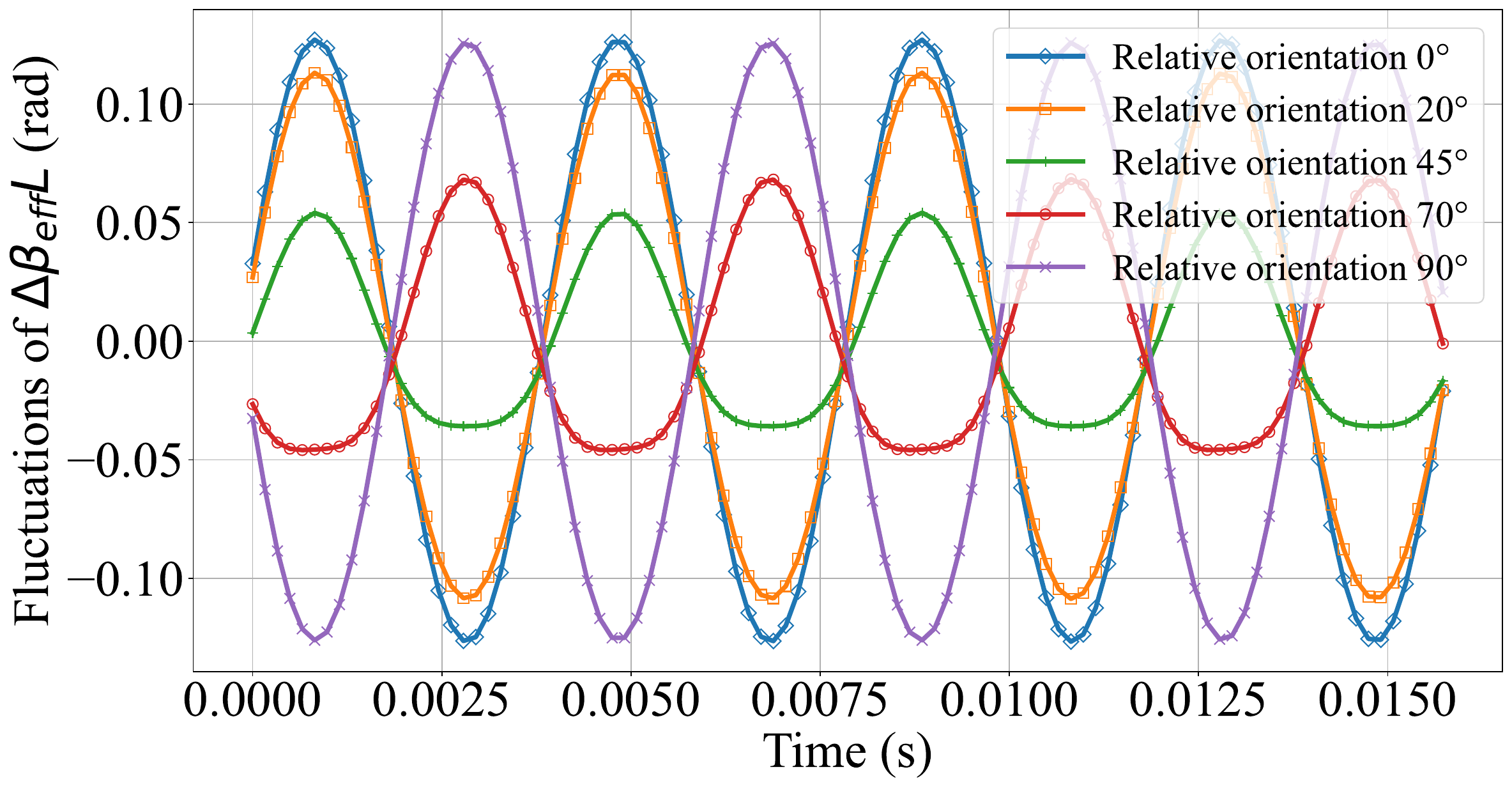}
    \caption{Fluctuations of estimated retardance $\Delta \beta_{eff}L$ for different relative orientations between fast axes of intrinsic birefringence and event-induced birefringence}
    \label{fig:diff_orientations}
\end{figure}

While the proposed method allows to infer some information on the nature of the deformation, it does not provide quantification. Indeed, as discussed in section \ref{sec:transverse strain}, the estimated effective birefringence does not vary linearly with the amplitude of the perturbation. Its response depends on the relative orientation between the intrinsic and external birefringence. We illustrate this through Fig. \ref{fig:diff_orientations} that shows the estimated retardance $\Delta \beta_{eff}L$ for an applied transverse strain (a sine wave of amplitude $100$ n$\varepsilon$ on $\varepsilon_y$ with an offset of $50$ n$\varepsilon$, $\varepsilon_x = -2\varepsilon_y$, and frequency $250$ Hz), for different angles of application with respect to the fast axis of the intrinsic birefringence of the affected region. The simulation was performed without the presence of noise in this case, to better see the effect of the angle between the induced birefringence orientation and the intrinsic birefringence orientation. Unless the fast axes of external and intrinsic birefringence are aligned (relative orientation $0$ rad) or orthogonal (relative orientation $\frac{\pi}{2}$ rad), the response is nonlinear in the perturbation, which is in agreement with Equation~\eqref{resulting_biref}.
In conclusion, the estimated effective birefringence strength being nonlinear in the perturbation depending on the strain orientation with respect to the fiber segment's fast axis makes it complicated to precisely quantify the  measurements. 
Retrieving further information on the effective birefringence orientation and not only on its magnitude could help separating external from internal contributions and better characterize the perturbations. 

Other limitations should be noted. One major concern for deployment in real-life infrastructures is the transfer of transverse strains from the cable to the fiber, which is expected to be very low \cite{kuvshinov_interaction_2016}, meaning only large dynamic transverse events might be actually detectable. Additionally, practical environmental disturbances are often complex, combining both longitudinal and transverse components. Deeper understanding of the different cable geometries and of coupling mechanisms is required to assess the efficiency of our proposed system in real-life applications.

\section{Conclusion}
In this work, we investigated the possibility to exploit both phase and polarization estimates in a coded MIMO-DAS architecture to better detect and discriminate events.
First, we developed a model for the MIMO-DAS taking into account common phase and polarization parameters. We model the scatterers spatial distribution along the fiber length and factor in the intrinsic linear birefringence in static conditions, and propose a way to integrate dynamic perturbations of both phase and birefringence in the described Jones waveplate model. For validation, we consider two types of events: pure longitudinal strain and anisotropic transverse strain, and show the ability of the proposed system to detect and localize both events with a mean spatial resolution of approximately $1.3$m, both in simulation and lab experiments. Moreover, we show that the estimated quantities exhibit different responses depending on the type of strain. While the differential phase is sensitive to both longitudinal and transverse strain, the estimated effective linear birefringence strength is mostly impacted by transverse strain only, allowing for discrimination between the two events. Future perspectives would include a better understanding on the effect of noise on the estimated quantities, and experimental demonstrations in field trials. 

\section*{Acknowledgments}

This work received funding from the AID (Agence de l’Innovation de Défense) in the frame of the FIBROUS project.

\bibliography{Biblio}

@inproceedings{cedilnik_advances_2018,
	address = {Lausanne},
	title = {Advances in {Train} and {Rail} {Monitoring} with {DAS}},
	isbn = {978-1-943580-50-7},
	doi = {10.1364/OFS.2018.ThE35},
	language = {en},
	booktitle = {26th {International} {Conference} on {Optical} {Fiber} {Sensors}},
	publisher = {OSA},
	author = {Cedilnik, Gregor and Hunt, Ryan and Lees, Gareth},
	year = {2018},
	pages = {ThE35},
}

@article{dorize_enhancing_2018,
	title = {Enhancing performance of coherent {OTDR} systems with polarization diversity complementary codes},
	volume = {26},
	issn = {1094-4087},
	doi = {10.1364/OE.26.012878},
	number = {10},
	journal = {Optics Express},
	author = {Dorize, Christian and Awwad, Elie},
	month = may,
	year = {2018},
	note = {arXiv:1802.06641 [eess]},
	pages = {12878},
}

@article{guerrier_introducing_2020,
	title = {Introducing {Coherent} {MIMO} {Sensing}, a fading-resilient, polarization-independent approach to phase-{OTDR}},
	volume = {28},
	issn = {1094-4087},
	doi = {10.1364/OE.396460},
	number = {14},
	journal = {Optics Express},
	author = {Guerrier, Sterenn and Dorize, Christian and Awwad, Elie and Renaudier, Jérémie},
	month = jul,
	year = {2020},
	note = {arXiv:2005.07135 [eess]},
	pages = {21081},
}

@book{hartog_introduction_2017,
	edition = {1},
	title = {An {Introduction} to {Distributed} {Optical} {Fibre} {Sensors}},
	isbn = {978-1-315-11901-4},
	language = {en},
	publisher = {CRC Press},
	author = {Hartog, Arthur H.},
	month = may,
	year = {2017},
	doi = {10.1201/9781315119014},
}

@article{lindsey_fiberoptic_2017,
	title = {Fiber‐{Optic} {Network} {Observations} of {Earthquake} {Wavefields}},
	volume = {44},
	copyright = {http://onlinelibrary.wiley.com/termsAndConditions\#am},
	issn = {0094-8276, 1944-8007},
	doi = {10.1002/2017GL075722},
	language = {en},
	number = {23},
	journal = {Geophysical Research Letters},
	author = {Lindsey, Nathaniel J. and Martin, Eileen R. and Dreger, Douglas S. and Freifeld, Barry and Cole, Stephen and James, Stephanie R. and Biondi, Biondo L. and Ajo‐Franklin, Jonathan B.},
	month = dec,
	year = {2017},
}

@article{westbrook_enhanced_2023,
	title = {Enhanced {Backscatter} {Fibers} for {Sensing} in {Telecom} {Networks}},
	volume = {41},
	copyright = {https://creativecommons.org/licenses/by/4.0/legalcode},
	issn = {0733-8724, 1558-2213},
	doi = {10.1109/JLT.2022.3225750},
	number = {3},
	journal = {Journal of Lightwave Technology},
	author = {Westbrook, Paul S. and Kremp, Tristan and Zhu, Benyuan and Ko, Wing and Shi, Zhou and Feder, Kenneth S.},
	month = feb,
	year = {2023},
	pages = {1010--1016},
}

@phdthesis{guerrier_high_2022,
	type = {PhD thesis},
	title = {High bandwidth detection of mechanical stress in optical fibre using coherent detection of {Rayleigh} scattering},
	school = {Institut Polytechnique de Paris},
	author = {Guerrier, Sterenn},
	month = feb,
	year = {2022},
}

@article{palmieri_distributed_2013,
	title = {Distributed {Optical} {Fiber} {Sensing} {Based} on {Rayleigh} {Scattering}},
	volume = {7},
	issn = {18743285},
	doi = {10.2174/1874328501307010104},
	language = {en},
	number = {1},
	journal = {The Open Optics Journal},
	author = {Palmieri, Luca},
	month = dec,
	year = {2013},
	pages = {104--127},
}

@article{galtarossa_reflectometric_2008,
	title = {Reflectometric {Characterization} of {Hinges} in {Optical} {Fiber} {Links}},
	volume = {20},
	copyright = {https://ieeexplore.ieee.org/Xplorehelp/downloads/license-information/IEEE.html},
	issn = {1041-1135, 1941-0174},
	doi = {10.1109/LPT.2008.921845},
	language = {en},
	number = {10},
	journal = {IEEE Photonics Technology Letters},
	author = {Galtarossa, Andrea and Grosso, Daniele and Palmieri, Luca and Schenato, Luca},
	month = may,
	year = {2008},
	pages = {854--856},
}

@article{costa_localization_2023,
	title = {Localization of seismic waves with submarine fiber optics using polarization-only measurements},
	volume = {2},
	issn = {2731-3395},
	doi = {10.1038/s44172-023-00138-4},
	language = {en},
	number = {1},
	journal = {Communications Engineering},
	author = {Costa, Luis and Varughese, Siddharth and Mertz, Pierre and Kamalov, Valey and Zhan, Zhongwen},
	month = dec,
	year = {2023},
	pages = {86},
}

@inproceedings{yaman_polarization_2023,
	address = {San Diego, CA, USA},
	title = {Polarization {Sensing} {Using} {Polarization} {Rotation} {Matrix} {Eigenvalue} {Method}},
	copyright = {https://doi.org/10.15223/policy-029},
	doi = {10.23919/OFC49934.2023.10117264},
	language = {en},
	booktitle = {2023 {Optical} {Fiber} {Communications} {Conference} and {Exhibition} ({OFC})},
	publisher = {IEEE},
	author = {Yaman, Fatih and Li, Yang and Han, Shaobo and Inoue, Takanori and Mateo, Eduardo and Inada, Yoshihisa},
	month = mar,
	year = {2023},
	pages = {1--3},
}

@article{mecozzi_sensing_2024,
	title = {Sensing with submarine optical cables},
	volume = {9},
	issn = {2378-0967},
	doi = {10.1063/5.0210825},
	language = {en},
	number = {7},
	journal = {APL Photonics},
	author = {Mecozzi, Antonio},
	month = jul,
	year = {2024},
	pages = {070902},
}

@article{lu_distributed_2019,
	title = {Distributed optical fiber sensing: {Review} and perspective},
	volume = {6},
	issn = {1931-9401},
	shorttitle = {Distributed optical fiber sensing},
	doi = {10.1063/1.5113955},
	language = {en},
	number = {4},
	journal = {Applied Physics Reviews},
	author = {Lu, Ping and Lalam, Nageswara and Badar, Mudabbir and Liu, Bo and Chorpening, Benjamin T. and Buric, Michael P. and Ohodnicki, Paul R.},
	month = dec,
	year = {2019},
	pages = {041302},
}

@incollection{yasin_optical_2012,
	title = {Optical {Fiber} {Birefringence} {Effects} – {Sources}, {Utilization} and {Methods} of {Suppression}},
	isbn = {978-953-307-823-6},
	language = {en},
	booktitle = {Recent {Progress} in {Optical} {Fiber} {Research}},
	publisher = {InTech},
	author = {Drexler, Petr and Fial, Pavel},
	editor = {Yasin, Moh.},
	month = jan,
	year = {2012},
	doi = {10.5772/27517},
}

@book{sharpe_springer_2008,
	address = {Boston, MA},
	series = {{SpringerLink} {Bücher}},
	title = {Springer {Handbook} of {Experimental} {Solid} {Mechanics}},
	isbn = {978-0-387-26883-5 978-0-387-30877-7},
	language = {en},
	publisher = {Springer Science+Business Media},
	author = {Sharpe, William N.},
	year = {2008},
	doi = {10.1007/978-0-387-30877-7},
}

@article{ulrich_bending-induced_1980,
	title = {Bending-induced birefringence in single-mode fibers},
	volume = {5},
	copyright = {https://doi.org/10.1364/OA\_License\_v1\#VOR},
	issn = {0146-9592, 1539-4794},
	doi = {10.1364/OL.5.000273},
	language = {en},
	number = {6},
	journal = {Optics Letters},
	author = {Ulrich, R. and Rashleigh, S. C. and Eickhoff, W.},
	month = jun,
	year = {1980},
	pages = {273},
}

@article{wuilpart_measurement_2001,
	title = {Measurement of the spatial distribution of birefringence in optical fibers},
	volume = {13},
	copyright = {https://ieeexplore.ieee.org/Xplorehelp/downloads/license-information/IEEE.html},
	issn = {1041-1135, 1941-0174},
	doi = {10.1109/68.935820},
	language = {en},
	number = {8},
	journal = {IEEE Photonics Technology Letters},
	author = {Wuilpart, M. and Megret, P. and Blondel, M. and Rogers, A.J. and Defosse, Y.},
	month = aug,
	year = {2001},
	pages = {836--838},
}

@article{rogers_distributed_2000,
	title = {Distributed {Measurement} of {Strain} using {Optical}‐fibre {Backscatter} {Polarimetry}},
	volume = {36},
	copyright = {http://onlinelibrary.wiley.com/termsAndConditions\#vor},
	issn = {0039-2103, 1475-1305},
	doi = {10.1111/j.1475-1305.2000.tb01189.x},
	language = {en},
	number = {3},
	journal = {Strain},
	author = {Rogers, A.J.},
	month = aug,
	year = {2000},
	pages = {135--142},
}

@book{huard_polarisation_1994,
	address = {Paris},
	title = {Polarisation de la lumière},
	isbn = {978-2-225-84300-6},
	language = {fre},
	publisher = {Masson},
	author = {Huard, Serge},
	year = {1994},
	note = {OCLC: 1408598136},
}

@article{feng_distributed_2018,
	title = {Distributed polarization analysis with binary polarization rotators for the accurate measurement of distance-resolved birefringence along a single-mode fiber},
	volume = {26},
	copyright = {https://doi.org/10.1364/OA\_License\_v1\#VOR-OA},
	issn = {1094-4087},
	doi = {10.1364/oe.26.025989},
	language = {en},
	number = {20},
	journal = {Optics Express},
	author = {Feng, Ting and Shang, Yanling and Wang, Xichen and Wu, Shengbao and Khomenko, Anton and Chen, Xiaojun and Yao, X. Steve},
	month = oct,
	year = {2018},
	note = {Publisher: Optica Publishing Group},
	pages = {25989},
}

@article{chen_distributed_2021,
	title = {Distributed {Fiber} {Birefringence} {Measurement} {Using} {Pulse}-{Compression} $\phi$-{OTDR}},
	volume = {11},
	issn = {1674-9251, 2190-7439},
	doi = {10.1007/s13320-020-0604-3},
	language = {en},
	number = {4},
	journal = {Photonic Sensors},
	author = {Chen, Yongxiang and Fu, Yun and Xiong, Ji and Wang, Zinan},
	month = dec,
	year = {2021},
	pages = {402--410},
}

@book{pachnicke_fiber-optic_2012,
	address = {Berlin, Heidelberg},
	series = {Signals and {Communication} {Technology}},
	title = {Fiber-{Optic} {Transmission} {Networks}: {Efficient} {Design} and {Dynamic} {Operation}},
	copyright = {https://www.springernature.com/gp/researchers/text-and-data-mining},
	isbn = {978-3-642-21054-9 978-3-642-21055-6},
	shorttitle = {Fiber-{Optic} {Transmission} {Networks}},
	language = {en},
	publisher = {Springer Berlin Heidelberg},
	author = {Pachnicke, Stephan},
	year = {2012},
	doi = {10.1007/978-3-642-21055-6},
}

@article{zhao_nonlocal_2025,
	title = {Nonlocal {Polarization} {Effect} in {Phase}-{Sensitive} {Optical} {Time}-{Domain} {Reflectometry}},
	volume = {43},
	copyright = {https://ieeexplore.ieee.org/Xplorehelp/downloads/license-information/IEEE.html},
	issn = {0733-8724, 1558-2213},
	doi = {10.1109/JLT.2025.3580816},
	language = {en},
	number = {16},
	journal = {Journal of Lightwave Technology},
	author = {Zhao, Can and Zhang, Youmin and Hu, Zihe and Tang, Ming},
	month = aug,
	year = {2025},
	pages = {7981--7990},
}

@book{damask_polarization_2005,
	address = {New York, NY},
	series = {Springer {Series} in {Optical} {Sciences}},
	title = {Polarization {Optics} in {Telecommunications}},
	isbn = {978-0-387-22493-0 978-0-387-26302-1},
	language = {en},
	number = {101},
	publisher = {Springer New York},
	author = {Damask, Jay N.},
	year = {2005},
	doi = {10.1007/b137386},
}

@article{palmieri_distributed_2013-polar,
	title = {Distributed polarimetric measurements for optical fiber sensing},
	volume = {19},
	issn = {10685200},
	doi = {10.1016/j.yofte.2013.07.015},
	language = {en},
	number = {6},
	journal = {Optical Fiber Technology},
	author = {Palmieri, Luca},
	month = dec,
	year = {2013},
	pages = {720--728},
}

@article{masoudi_analysis_2017,
	title = {Analysis of distributed optical fibre acoustic sensors through numerical modelling},
	volume = {25},
	issn = {1094-4087},
	doi = {10.1364/OE.25.032021},
	language = {en},
	number = {25},
	journal = {Optics Express},
	author = {Masoudi, Ali and Newson, Trevor P.},
	month = dec,
	year = {2017},
	pages = {32021},
}

@article{liokumovich_fundamentals_2015,
	title = {Fundamentals of {Optical} {Fiber} {Sensing} {Schemes} {Based} on {Coherent} {Optical} {Time} {Domain} {Reflectometry}: {Signal} {Model} {Under} {Static} {Fiber} {Conditions}},
	volume = {33},
	copyright = {https://ieeexplore.ieee.org/Xplorehelp/downloads/license-information/IEEE.html},
	issn = {0733-8724, 1558-2213},
	shorttitle = {Fundamentals of {Optical} {Fiber} {Sensing} {Schemes} {Based} on {Coherent} {Optical} {Time} {Domain} {Reflectometry}},
	doi = {10.1109/JLT.2015.2449085},
	language = {en},
	number = {17},
	journal = {Journal of Lightwave Technology},
	author = {Liokumovich, Leonid B. and Ushakov, Nikolai A. and Kotov, Oleg I. and Bisyarin, Mikhail A. and Hartog, Arthur H.},
	month = sep,
	year = {2015},
	pages = {3660--3671},
}

@article{gafsi_analysis_2000,
	title = {Analysis of {Induced}-{Birefringence} {Effects} on {Fiber} {Bragg} {Gratings}},
	volume = {6},
	copyright = {https://www.elsevier.com/tdm/userlicense/1.0/},
	issn = {10685200},
	doi = {10.1006/ofte.2000.0333},
	language = {en},
	number = {3},
	journal = {Optical Fiber Technology},
	author = {Gafsi, Rachid and El-Sherif, Mahmoud A.},
	month = jul,
	year = {2000},
	pages = {299--323},
}

@article{kuvshinov_interaction_2016,
	title = {Interaction of helically wound fibre‐optic cables with plane seismic waves},
	volume = {64},
	issn = {0016-8025, 1365-2478},
	doi = {10.1111/1365-2478.12303},
	language = {en},
	number = {3},
	journal = {Geophysical Prospecting},
	author = {Kuvshinov, B.N.},
	month = may,
	year = {2016},
	pages = {671--688},
}

@article{wai_polarization_1994,
	title = {Polarization decorrelation in optical fibers with randomly varying birefringence},
	volume = {19},
	copyright = {https://doi.org/10.1364/OA\_License\_v1\#VOR},
	issn = {0146-9592, 1539-4794},
	doi = {10.1364/OL.19.001517},
	language = {en},
	number = {19},
	journal = {Optics Letters},
	author = {Wai, P. K. A. and Menyuk, C. R.},
	month = oct,
	year = {1994},
	pages = {1517},
}

@article{wai_polarization_1996,
	title = {Polarization mode dispersion, decorrelation, and diffusion in optical fibers with randomly varying birefringence},
	volume = {14},
	copyright = {https://ieeexplore.ieee.org/Xplorehelp/downloads/license-information/IEEE.html},
	issn = {0733-8724, 1558-2213},
	doi = {10.1109/50.482256},
	language = {en},
	number = {2},
	journal = {Journal of Lightwave Technology},
	author = {Wai, P.K.A. and Menyuk, C.R.},
	month = feb,
	year = {1996},
	pages = {148--157},
}

@article{corsi_beat_1999,
	title = {Beat length characterization based on backscattering analysis in randomly perturbed single-mode fibers},
	volume = {17},
	copyright = {https://ieeexplore.ieee.org/Xplorehelp/downloads/license-information/IEEE.html},
	issn = {07338724},
	doi = {10.1109/50.774250},
	language = {en},
	number = {7},
	journal = {Journal of Lightwave Technology},
	author = {Corsi, F. and Galtarossa, A. and Palmieri, L.},
	month = jul,
	year = {1999},
	pages = {1172--1178},
}

@article{tu_theoretical_2025,
	title = {A theoretical model for small-area transverse force measurement based on linearly chirped fiber {Bragg} grating},
	volume = {131},
	issn = {0946-2171, 1432-0649},
	doi = {10.1007/s00340-025-08439-6},
	language = {en},
	number = {4},
	journal = {Applied Physics B},
	author = {Tu, Xinghua and Ou, Jie and Li, Lianyan and Zhao, Haiyang and Diao, Junhui},
	month = apr,
	year = {2025},
	pages = {81},
}

@article{Bertholds_86,
    author = {A. Bertholds and R. D\"{a}ndliker},
    journal = {Appl. Opt.},
    number = {3},
    pages = {340--343},
    publisher = {Optica Publishing Group},
    title = {High-resolution photoelastic pressure sensor using low-birefringence fiber},
    volume = {25},
    month = {Feb},
    year = {1986},
    doi = {10.1364/AO.25.000340},
}

@article{SPIEPaper,
author = {D. Prato and M. Mokhtari Sheramin and R. Gabet and E. Awwad},
journal = {accepted to SPIE Photonics Europe},
title = {Event Detection and Localization Using a Multiple-Input-Multiple-Output Distributed Fiber Sensor with Birefringence and Phase Estimation},
year={2026}
}
\bibliographystyle{IEEEtran}
\clearpage 

\end{document}